\documentclass[12pt, journal, onecolumn]{IEEEtran}

\usepackage{graphicx, subcaption}
\graphicspath{{figs2/}} 

\usepackage[nowarn,acronyms,nonumberlist,nopostdot,nomain,nogroupskip]{glossaries}
\usepackage{multirow}
\usepackage{setspace}
\usepackage{tikz}
\usepackage{lipsum}

\newcommand\copyrighttext{%
  \centering\footnotesize This work has been submitted to the IEEE for possible publication.  Copyright may be transferred without notice, after which this version may no longer be accessible.}
\newcommand\copyrightnotice{%
\begin{tikzpicture}[remember picture,overlay]
\node[anchor=north,yshift=-5pt] at (current page.north) {\fbox{\parbox{\dimexpr\textwidth-\fboxsep-\fboxrule\relax}{\copyrighttext}}};
\end{tikzpicture}%
}

\usepackage{optidef}
\usepackage{amssymb}
\usepackage{mathtools}

\usepackage{algorithm}
\usepackage{algpseudocode}

\newacronym{3gpp}{3GPP}{3rd Generation Partnership Project}
\newacronym{5g}{5G}{5\textsuperscript{th} Generation}
\newacronym{6g}{6G}{6\textsuperscript{th} Generation}
\newacronym{5gc}{5GC}{5G Core}
\newacronym{bs}{BS}{Base Station}
\newacronym{abft}{A-BFT}{Association-BeamForming Training}
\newacronym[firstplural=Access Categories (ACs)]{ac}{AC}{Access Category}
\newacronym{adc}{ADC}{Analog to Digital Converter}
\newacronym{addts}{ADDTS}{Add Traffic Stream}
\newacronym{afbw}{AFBW}{Average Fading Bandwidth}
\newacronym{aid}{AID}{Association Identifier}
\newacronym{aimd}{AIMD}{Additive Increase Multiplicative Decrease}
\newacronym{am}{AM}{Acknowledged Mode}
\newacronym{amc}{AMC}{Adaptive Modulation and Coding}
\newacronym{ampdu}{A-MPDU}{MAC Protocol Data Unit Aggregation}
\newacronym{mmse}{MMSE}{Minimum Mean Square Error}
\newacronym{aoa}{AoA}{Angle of Arrival}
\newacronym{aod}{AoD}{Angle of Departure}
\newacronym{ap}{AP}{Access Point}
\newacronym{app}{APP}{Application Layer}
\newacronym{aqm}{AQM}{Active Queue Management}
\newacronym{ar}{AR}{Augmented Reality}
\newacronym{ati}{ATI}{Announcement Transmission Interval}
\newacronym{awgn}{AGWN}{Additive White Gaussian Noise}
\newacronym{awv}{AWV}{Antenna Weight Vector}
\newacronym{b5g}{B5G}{beyond-5G}
\newacronym{balia}{BALIA}{Balanced Link Adaptation}
\newacronym{bdp}{BDP}{Bandwidth-Delay Product}
\newacronym{bf}{BF}{Beamforming}
\newacronym{bhi}{BHI}{Beacon Header Interval}
\newacronym{bi}{BI}{Beacon Interval}
\newacronym{brp}{BRP}{Beam Refinement Protocol}
\newacronym{bss}{BSS}{Basic Service Set}
\newacronym{bti}{BTI}{Beacon Transmission Interval}
\newacronym{cad}{CAD}{Computer-aided Design}
\newacronym{cbap}{CBAP}{Contention-Based Access Period}
\newacronym{cbr}{CBR}{Constant Bitrate}
\newacronym{cc}{CC}{Congestion Control}
\newacronym{cdf}{CDF}{Cumulative Distribution Function}
\newacronym{cf}{CF}{Cell-free}
\newacronym{cfmmimo}{CF-mMIMO}{Cell-Free massive Multiple-Input, Multiple-Output}
\newacronym{cir}{CIR}{Channel Impulse Response}
\newacronym{cn}{CN}{Core Network}
\newacronym{comp}{CoMP}{Coordinated MultiPoint}
\newacronym{cp}{CP}{Control Plane}
\newacronym{cpu}{CPU}{central processing unit}
\newacronym{cqi}{CQI}{Channel Quality Indicator}
\newacronym{cran}{C-RAN}{cloud radio access network}
\newacronym{crs}{CRS}{Cell Reference Signal}
\newacronym{csi}{CSI}{Channel State Information}
\newacronym{csirs}{CSI-RS}{Channel State Information - Reference Signal}
\newacronym{csmaca}{CSMA/CA}{Carrier Sense Multiple Access with Collision Avoidance}
\newacronym{cts}{CTS}{Clear to Send}
\newacronym{d2d}{D2D}{Device-to-device}
\newacronym{dcp}{DCP}{Diverse Clustering Problem}
\newacronym{dcc}{DCC}{Diverse Capacitated Clustering}
\newacronym{dce}{DCE}{Direct Code Execution}
\newacronym{dcf}{DCF}{Distributed Coordination Function}
\newacronym{dci}{DCI}{Downlink Control Information}
\newacronym{delts}{DELTS}{Delete Traffic Stream}
\newacronym{dked}{DKED}{Double Knife Edge Diffraction}
\newacronym{dl}{DL}{Downlink}
\newacronym{dmg}{DMG}{Directional Multi-Gigabit}
\newacronym{dmr}{DMR}{Deadline Miss Ratio}
\newacronym{dmrs}{DMRS}{DeModulation Reference Signal}
\newacronym{dpp}{DPP}{Determinantal Point Processes}
\newacronym{dti}{DTI}{Data Transmission Interval}
\newacronym{dtmke}{DTMKE}{Double-truncated Multiple Knife-edge}
\newacronym{e2e}{E2E}{End-to-End}
\newacronym{ecn}{ECN}{Explicit Congestion Notification}
\newacronym{edca}{EDCA}{Enhanced Distributed Channel Access}
\newacronym{edf}{EDF}{Earliest Deadline First}
\newacronym{enb}{eNB}{evolved Node Base}
\newacronym{endc}{EN-DC}{E-UTRAN-\gls{nr} \gls{dc}}
\newacronym{epc}{EPC}{Evolved Packet Core}
\newacronym{es}{ES}{equal size}
\newacronym{ese}{ESE}{Extended Schedule Element}
\newacronym{fdd}{FDD}{Frequency Division Duplexing}
\newacronym{fdma}{FDMA}{Frequency Division Multiple Access}
\newacronym{fov}{FoV}{Field-of-View}
\newacronym{fs}{FS}{Fast Switching}
\newacronym{ftp}{FTP}{File Transfer Protocol}
\newacronym{gnb}{gNB}{Next Generation Node Base}
\newacronym{harq}{HARQ}{Hybrid Automatic Repeat reQuest}
\newacronym{hetnet}{HetNet}{Heterogeneous Network}
\newacronym{hh}{HH}{Hard Handover}
\newacronym{hol}{HOL}{Head-of-Line}
\newacronym{hqf}{HQF}{Highest-quality-first}
\newacronym{ia}{IA}{Initial Access}
\newacronym{iab}{IAB}{Integrated Access and Backhaul}
\newacronym{ibss}{IBSS}{Independent Basic Service Set}
\newacronym{id}{ID}{Identifier}
\newacronym{imt}{IMT}{International Mobile Telecommunication}
\newacronym{inr}{INR}{Interference to Noise Ratio}
\newacronym{iot}{IoT}{Internet of Things}
\newacronym{ipa}{IPA}{Inter-Packet Arrival}
\newacronym{ism}{ISM}{Industrial, Scientific, and Medical}
\newacronym{kpi}{KPI}{Key Performance Indicator}
\newacronym{lcf}{LCF}{Level Crossing Frequency}
\newacronym{lcr}{LCR}{Level Crossing Rate}
\newacronym{los}{LoS}{Line-of-Sight}
\newacronym{lp}{LP}{Low Power}
\newacronym{lsf}{LSF}{large-scale fading}
\newacronym{lte}{LTE}{Long Term Evolution}
\newacronym{m2m}{M2M}{Machine to Machine}
\newacronym{mac}{MAC}{Medium Access Control}
\newacronym{mc}{MC}{Multi-Connectivity}
\newacronym{mcs}{MCS}{Modulation and Coding Scheme}
\newacronym{mdgp}{MDGP}{Maximally Diverse Grouping Problem}
\newacronym{mec}{MEC}{Mobile Edge Cloud}
\newacronym{mi}{MI}{Mutual Information}
\newacronym{mib}{MIB}{Master Information Block}
\newacronym{mimo}{MIMO}{Multiple Input, Multiple Output}
\newacronym{mmimo}{mMIMO}{massive Multiple-Input, Multiple-Output}
\newacronym{mumimo}{MU-MIMO}{Multi-User Multiple Input, Multiple Output}
\newacronym{ml}{ML}{Machine Learning}
\newacronym{mlr}{MLR}{Maximum-local-rate}
\newacronym[plural=\gls{mme}s,firstplural=Mobility Management Entities (MMEs)]{mme}{MME}{Mobility Management Entity}
\newacronym{mmwave}{mmWave}{Millimeter Wave}
\newacronym{moi}{MoI}{Method of Images}
\newacronym{mpc}{MPC}{Multi Path Component}
\newacronym{mptcp}{MPTCP}{Multipath TCP}
\newacronym{mr}{MR}{Maximal Ratio}
\newacronym{mrdc}{MR-DC}{Multi \gls{rat} \gls{dc}}
\newacronym{mss}{MSS}{Maximum Segment Size}
\newacronym{mtd}{MTD}{Machine-Type Device}
\newacronym{mtu}{MTU}{Maximum Transmission Unit}
\newacronym{nav}{NAV}{Network Allocation Vector}
\newacronym{ncbr}{NCBR}{Non-Constant Bitrate}
\newacronym{nfv}{NFV}{Network Function Virtualization}
\newacronym{nlos}{NLoS}{Non-Line-of-Sight}
\newacronym{nr}{NR}{New Radio}
\newacronym{nrmse}{NRMSE}{Normalized Root Mean Square Error}
\newacronym{ns3}{ns-3}{Network Simulator 3}
\newacronym{nsa}{NSA}{Non Stand Alone}
\newacronym{o2i}{O2I}{Outdoor-to-Indoor}
\newacronym{ofdm}{OFDM}{Orthogonal Frequency Division Multiplexing}
\newacronym{pa}{PA}{Position-aware}
\newacronym{pan}{PAN}{Personal Area Network}
\newacronym{pas}{PAS}{Power Angular Spectrum}
\newacronym{pbch}{PBCH}{Physical Broadcast Channel}
\newacronym{pbss}{PBSS}{Personal Basic Service Set}
\newacronym{pci}{PCI}{Physical Cell Identity}
\newacronym{pcp}{PCP}{\gls{pbss} Central Point}
\newacronym{pcpap}{PCP/AP}{\acrlong{pcp}/\acrlong{ap}}
\newacronym{pdcch}{PDCCH}{Physical Downlonk Control Channel}
\newacronym{pdcp}{PDCP}{Packet Data Convergence Protocol}
\newacronym{pdsch}{PDSCH}{Physical Downlink Shared Channel}
\newacronym{pdu}{PDU}{Packet Data Unit}
\newacronym{pf}{PF}{Proportional Fair}
\newacronym{pgw}{PGW}{Packet Gateway}
\newacronym{phy}{PHY}{Physical Layer}
\newacronym{ppp}{PPP}{Poisson Point Process}
\newacronym{prb}{PRB}{Physical Resource Block}
\newacronym{pss}{PSS}{Primary Synchronization Signal}
\newacronym{pucch}{PUCCH}{Physical Uplink Control Channel}
\newacronym{pusch}{PUSCH}{Physical Uplink Shared Channel}
\newacronym{qd}{QD}{Quasi Deterministic}
\newacronym{qoe}{QoE}{Quality of Experience}
\newacronym{qos}{QoS}{Quality of Service}
\newacronym{rach}{RACH}{Random Access Channel}
\newacronym{ran}{RAN}{Radio Access Network}
\newacronym[firstplural=Radio Access Technologies (RATs)]{rat}{RAT}{Radio Access Technology}
\newacronym{red}{RED}{Random Early Detection}
\newacronym{rf}{RF}{Radio Frequency}
\newacronym{rl}{RL}{Reinforcement Learning}
\newacronym{rlc}{RLC}{Radio Link Control}
\newacronym{rlf}{RLF}{Radio Link Failure}
\newacronym{rms}{RMS}{Root Mean Square}
\newacronym{rr}{RR}{Round Robin}
\newacronym{rrc}{RRC}{Radio Resource Control}
\newacronym{rrm}{RRM}{Radio Resource Management}
\newacronym{rs}{RS}{Remote Server}
\newacronym{rsrp}{RSRP}{Reference Signal Received Power}
\newacronym{rsrq}{RSRQ}{Reference Signal Received Quality}
\newacronym{rss}{RSS}{Received Signal Strength}
\newacronym{rssi}{RSSI}{Received Signal Strength Indicator}
\newacronym{rt}{RT}{Ray Tracer}
\newacronym{rts}{RTS}{Request to Send}
\newacronym{rtt}{RTT}{Round Trip Time}
\newacronym{rw}{RW}{Receive Window}
\newacronym{rx}{RX}{Receiver}
\newacronym{sa}{SA}{standalone}
\newacronym{sack}{SACK}{Selective Acknowledgment}
\newacronym{sap}{SAP}{Service Access Point}
\newacronym{sc}{SC}{Single Carrier}
\newacronym{sch}{SCH}{Secondary Cell Handover}
\newacronym{scm}{SCM}{Spatial Channel Model}
\newacronym{scoot}{SCOOT}{Split Cycle Offset Optimization Technique}
\newacronym{sdma}{SDMA}{Spatial Division Multiple Access}
\newacronym{sdr}{SDR}{Software Defined Radio}
\newacronym{se}{SE}{Spectral Efficiency}
\newacronym{si}{SI}{Study Item}
\newacronym{sib}{SIB}{Secondary Information Block}
\newacronym{sinr}{SINR}{Signal-to-Interference-plus-Noise Ratio}
\newacronym{sir}{SIR}{Signal-to-Interference Ratio}
\newacronym{sls}{SLS}{Sector-Level Sweep}
\newacronym{sm}{SM}{Saturation Mode}
\newacronym{snr}{SNR}{Signal-to-Noise Ratio}
\newacronym{son}{SON}{Self-Organizing Network}
\newacronym{sp}{SP}{Service Period}
\newacronym{spr}{SPR}{Service Period Request}
\newacronym{srs}{SRS}{Sounding Reference Signal}
\newacronym{ss}{SS}{Synchronization Signal}
\newacronym{ssb}{SSB}{\gls{ss}}
\newacronym{sss}{SSS}{Secondary Synchronization Signal}
\newacronym{ssw}{SSW}{Sector Sweep}
\newacronym{sta}{STA}{Station}
\newacronym{stb}{STB}{Set Top Box}
\newacronym{tb}{TB}{Transport Block}
\newacronym{tcp}{TCP}{Transmission Control Protocol}
\newacronym{tdd}{TDD}{Time Division Duplexing}
\newacronym{tdma}{TDMA}{Time Division Multiple Access}
\newacronym{tfl}{TfL}{Transport for London}
\newacronym{tgad}{TGad}{Task Group ad}
\newacronym{tgay}{TGay}{Task Group ay}
\newacronym{tsconst}{TSCONST}{Traffic Scheduling Constraint}
\newacronym{tm}{TM}{Transparent Mode}
\newacronym{trp}{TRP}{Transmitter Receiver Pair}
\newacronym{ts}{TS}{Traffic Stream}
\newacronym{tspec}{TSPEC}{Traffic Specification}
\newacronym{tti}{TTI}{Transmission Time Interval}
\newacronym{ttt}{TTT}{Time-to-Trigger}
\newacronym{tx}{TX}{Transmitter}
\newacronym[firstplural=Transmission Opportunities (TXOPs)]{txop}{TXOP}{Transmission Opportunity}
\newacronym{udp}{UDP}{User Datagram Protocol}
\newacronym{ue}{UE}{User Equipment}
\newacronym{ul}{UL}{Uplink}
\newacronym{um}{UM}{Unacknowledged Mode}
\newacronym{uma}{UMa}{Urban Macro}
\newacronym{uml}{UML}{Unified Modeling Language}
\newacronym{utc}{UTC}{Urban Traffic Control}
\newacronym{v2v}{V2V}{Vehicle-to-Vehicle}
\newacronym{vbr}{VBR}{Variable Bit Rate}
\newacronym{vm}{VM}{Virtual Machine}
\newacronym{vs}{VS}{variable size}
\newacronym{vr}{VR}{Virtual Reality}
\newacronym{wbf}{WBF}{Wired Bias Function}
\newacronym{wf}{WF}{Wired-first}
\newacronym{wifi}{Wi-Fi}{Wireless Fidelity}
\newacronym{wigig}{WiGig}{Wireless Gigabit}
\newacronym{wlan}{WLAN}{Wireless Local Area Network}
\newacronym{ber}{BER}{Bit Error Rate}
\newacronym{arf}{ARF}{Auto Rate Fallback}
\newacronym{semm}{SEMM}{SPCA-EDCA Mixed Mode}
\newacronym{ppdu}{PPDU}{PHY Protocol Data Unit}
\newacronym{udn}{UDN}{Ultra Dense Network}
\newacronym{dnn}{DNN}{Deep Neural Network}

\usepackage{xcolor}

\usepackage{bbold}
\usepackage{cleveref}
\Crefformat{figure}{#2Fig.~#1#3}
\Crefformat{table}{#2Tab.~#1#3}
\Crefformat{algorithm}{#2Alg.~#1#3}
\Crefformat{section}{#2Sec.~#1#3}
\Crefmultiformat{figure}{Figs.~#2#1#3}{ and~#2#1#3}{, #2#1#3}{ and~#2#1#3}

\makeglossaries
\glsdisablehyper
\usepackage{cite}

\begin{document}
\bstctlcite{IEEEexample:BSTcontrol}

\title{Pilot Reuse in Cell-Free Massive MIMO Systems: A Diverse Clustering Approach
\thanks{Salman Mohebi has received funding from the European Union’s Horizon 2020 research and innovation programme under the Marie Skłodowska-Curie Grant agreement No. 813999.}}
\author{Salman Mohebi, Andrea Zanella, Michele Zorzi\\
Department of Information Engineering, University of Padova, Padova, Italy \\
Emails: \texttt{\{surname\}@dei.unipd.it}}

\maketitle

\setstretch{1.95}

\begin{abstract}\label{sec:abstract}
Distributed or \gls{cf} \gls{mmimo}, has been recently proposed as an answer to the limitations of the current \textit{network-centric} systems in providing high-rate ubiquitous transmission.
The capability of providing uniform service level makes \gls{cf} \gls{mmimo} a potential technology for beyond-5G and 6G networks.
The acquisition of accurate \gls{csi} is critical for different \gls{cf} \gls{mmimo} operations.
Hence, an uplink pilot training phase is used to efficiently estimate transmission channels.
The number of available orthogonal pilot signals is limited, and reusing these pilots will increase co-pilot interference.
This causes an undesirable effect known as pilot contamination that could reduce the system performance.
Hence, a proper pilot reuse strategy is needed to mitigate the effects of pilot contamination.
In this paper, we formulate pilot assignment in \gls{cf} \gls{mmimo} as a diverse clustering problem and propose an iterative maxima search scheme to solve it.
In this approach, we first form the clusters of \glspl{ue} so that the intra-cluster diversity maximizes and then assign the same pilots for all \glspl{ue} in the same cluster.
The numerical results show the proposed techniques' superiority over other methods concerning the achieved uplink and downlink average and per-user data rate.
\end{abstract}

\IEEEpeerreviewmaketitle

\copyrightnotice
\section{Introduction}\label{sec:intro}
\glsresetall
Distributed or \gls{cf} \gls{mmimo}~\cite{ngo2015cell, ngo2017cell, zhang2019cell, demir2021foundations, interdonato2019ubiquitous,elhoushy2021cell, bjornson2020scalable} has been considered as one of the potential technologies for \gls{b5g} and 6G wireless networks thanks to its capability of providing relatively uniform service to the \glspl{ue} in the coverage area.
As the name implies, \gls{cf} systems, unlike traditional cellular networks, do not consider the concepts of cell or cell boundaries, where a \gls{bs} serves multiple \glspl{ue} within its cell coverage.
In \gls{cf} \gls{mmimo}, \glspl{ue} are jointly served by a relatively larger number of geographically distributed \glspl{ap} over the same time-frequency resource. 
\gls{cf} \gls{mmimo} lies at the intersection of different technologies like \gls{mmimo}~\cite{lu2014overview,marzetta2015massive, marzetta2016fundamentals, larsson2014massive}, \gls{comp}~\cite{sawahashi2010coordinated, irmer2011coordinated} and \gls{udn}~\cite{kamel2016ultra, chen2016user}, combing the best of each technology, while eliminating their deficiencies~\cite{demir2021foundations}.
\gls{cf} \gls{mmimo} adopts its physical layer from cellular \gls{mmimo}, which brings 10x \gls{se} improvement over legacy cellular networks~\cite{demir2021foundations}.
This gain comes from deploying a massive number of antennas at each \gls{bs}, that provide spatial multiplexing for many \glspl{ue} by digital beamforming.
\gls{cf} \gls{mmimo} is inherently a distributed implementation of co-located \gls{mmimo}, where densely developed \glspl{ap} provide service for a smaller number of \glspl{ue} with a coherent joint transmission and \gls{tdd} operation.
The procedure takes place in a \textit{user-centric} (as opposed to \textit{network-centric}) fashion, where the \glspl{ue} are surrounded and served by several \glspl{ap}.
The \glspl{ap} are connected to one or multiple \glspl{cpu}, i.e., edge-cloud processors or \gls{cran}~\cite{wu2015cloud} data centers~\cite{bjornson2019making}, by high-capacity fronthaul links where data precoding/decoding operations, synchronization, and other network management operations take place.
The acquisition of accurate \gls{csi} is critical for different \gls{cf} \gls{mmimo} operations. 
Hence, it employs \gls{ul} pilots to estimate the channel at the \gls{ap} while eliminating the \gls{dl} pilot training phase by considering channel reciprocity.
However, due to the limited number of channel uses in each coherence interval, only a limited number of orthogonal pilots are available, typically smaller than the number of \glspl{ue}.
The number of available pilots is independent of the number of \glspl{ue} and is limited due to the natural channel variation in the time and frequency domains~\cite{chen2021survey}.
Hence, reusing the same pilot for different \glspl{ue} is inevitable, which introduces undesirable effects known as pilot contamination: due to the co-pilot interference among \glspl{ue}, the fading channel at the \glspl{ap} can not be accurately estimated.

\subsection{Related Works}\label{sec:related_works}
Different pilot assignment strategies have already been proposed in the literature to solve this issue.
The most straightforward approach is \textit{random} pilot assignment~\cite{ngo2017cell}, where each \gls{ap} independently assigns a random pilot to its associated \glspl{ue}.
This is a fully distributed procedure and requires a minimum degree of centralization and knowledge of the pilots of other \glspl{ue}, but has the worst performance among different pilot reuse strategies.
Hence, a better pilot assignment policy needs to know other \glspl{ue}' (at least its neighbors) pilots to reduce the effects of pilot contamination.
A \textit{greedy} pilot assignment is proposed in~\cite{ngo2017cell}, that iteratively updates the pilot of the \gls{ue} with minimum rate.
A \textit{structured} pilot assignment scheme is proposed in~\cite{attarifar2018random}, that maximizes the minimum distance of the co-pilot \glspl{ue}.
The location information of the \glspl{ue} is utilized in \textit{location-based greedy} pilot assignment~\cite{zhang2018location} to improve the initial pilot assignment.
The pilot assignment is also considered as an interference management problem with multiple group-casting messages in~\cite{yu2022topological} and then solved by \textit{topological} pilot assignments where both known and unknown \gls{ue}/\gls{ap} connectivity patterns are considered.

Graph theory has also been used to model pilot assignment in \gls{cf} \gls{mmimo}, where \textit{graph coloring}~\cite{liu2020graph} and \textit{weight graphic} \cite{zeng2021pilot} schemes created and employed an interference graph to assign pilots to different \glspl{ue}.
The authors in~\cite{liu2019tabu, ding2021improved} used \textit{tabu search} pilot assignment in \gls{cf} \gls{mmimo}.
Pilot assignment can also be considered as a graph matching problem and then solved by the \textit{Hungarian} algorithm~\cite{buzzi2020pilot}.
A \textit{weighted count-based} pilot assignment is presented in~\cite{li2021pilot}, that uses the \glspl{ue} prior geographic information and pilot power to maximize the pilot reuse weighted distance.
A scalable pilot assignment scheme is presented in~\cite{sarker2021granting} to grant massive access in \gls{cf} \gls{mmimo}.
Another scalable pilot assignment algorithm based on \textit{deep learning} is presented in~\cite{li2021scalable}, that uses \glspl{ue} geographical locations as an input.
The authors in~\cite{raharya2020pursuit} presented a \textit{pursuit learning} approach for joint pilot allocation and \gls{ap} association.
A pilot assignment strategy based on \textit{quantum bacterial foraging optimization} is proposed in~\cite{zhu2021pilot}.
The pilot assignment is also considered as a balanced diverse clustering problem in~\cite{mohebi2022repulsive} and solved by a \textit{repulsive clustering} approach.

\subsection{Motivation and Contributions}
The co-pilot interference, in principle, is caused by pilot reuse for similar \glspl{ue}, i.e., in terms of geographical proximity and/or similar channel coefficient. 
Hence, an intelligent pilot assignment scheme could employ the similarity information to avoid pilot reuse for similar \glspl{ue}.
Motivated by this consideration, in this paper, we consider pilot assignment as a \gls{dcp}, which forms the clusters of \glspl{ue} with maximized intra-cluster diversity.
We then propose an iterative maxima search method to solve this problem.
These clusters are then used for pilot assignments.

The main contributions of this paper are summarized as follows:
\begin{itemize}
    \item We formulated the pilot assignment as a \gls{dcp}: a clustering problem with arbitrary capacity constraints on cluster size, aiming to maximize the intra-cluster heterogeneity and inter-cluster homogeneity.
    \item We propose an iterative maxima search approach composed of a local search and weak and robust perturbation procedures to sufficiently cover the search space and balance between quality and diversity of solutions.
    \item We evaluate and compare the performance of the proposed approach under different situations and scenarios with respect to average and per-user \gls{ul} and \gls{dl} rates.
\end{itemize}

\subsection{Paper Outline and Notation}
The remainder of this paper is summarized as follows.
\Cref{sec:sysmodel} provides the system model for the \gls{cf} \gls{mmimo}. We formulate the pilot assignment problem in~\Cref{sec:problem_formulation} and then propose the iterative maxima search procedure in~\Cref{sec:iterative_optima_search}.
The numerical results are presented in \Cref{sec:result}, and finally, \Cref{sec:conclusion} concludes the paper.

\subsubsection*{Notation} Boldface letters denote column vectors.
The superscripts $(\cdot)^{*}$, $(\cdot)^{T}$, and $(\cdot)^{H}$ indicate the conjugate, transpose, and conjugate-transpose, respectively. 
The Euclidean norm and the expectation operators are denoted by $\Vert\cdot\Vert$ and $\mathbb{E}\{\cdot\}$, respectively. 
A capital calligraphic letter specifies a set.
Finally, $z\sim\mathcal{CN}(0,\sigma^2)$ represents a circularly symmetric complex Gaussian random variable $z$ with zero mean and variance $\sigma^2$, and $z\sim\mathcal{N}(0, \sigma^2)$ denotes a real valued Gaussian random variable.



\section{System Model}\label{sec:sysmodel}
\begin{figure}[t]
    \begin{subfigure}[b]{\linewidth}
         \centering
        \includegraphics[width=0.65\textwidth]{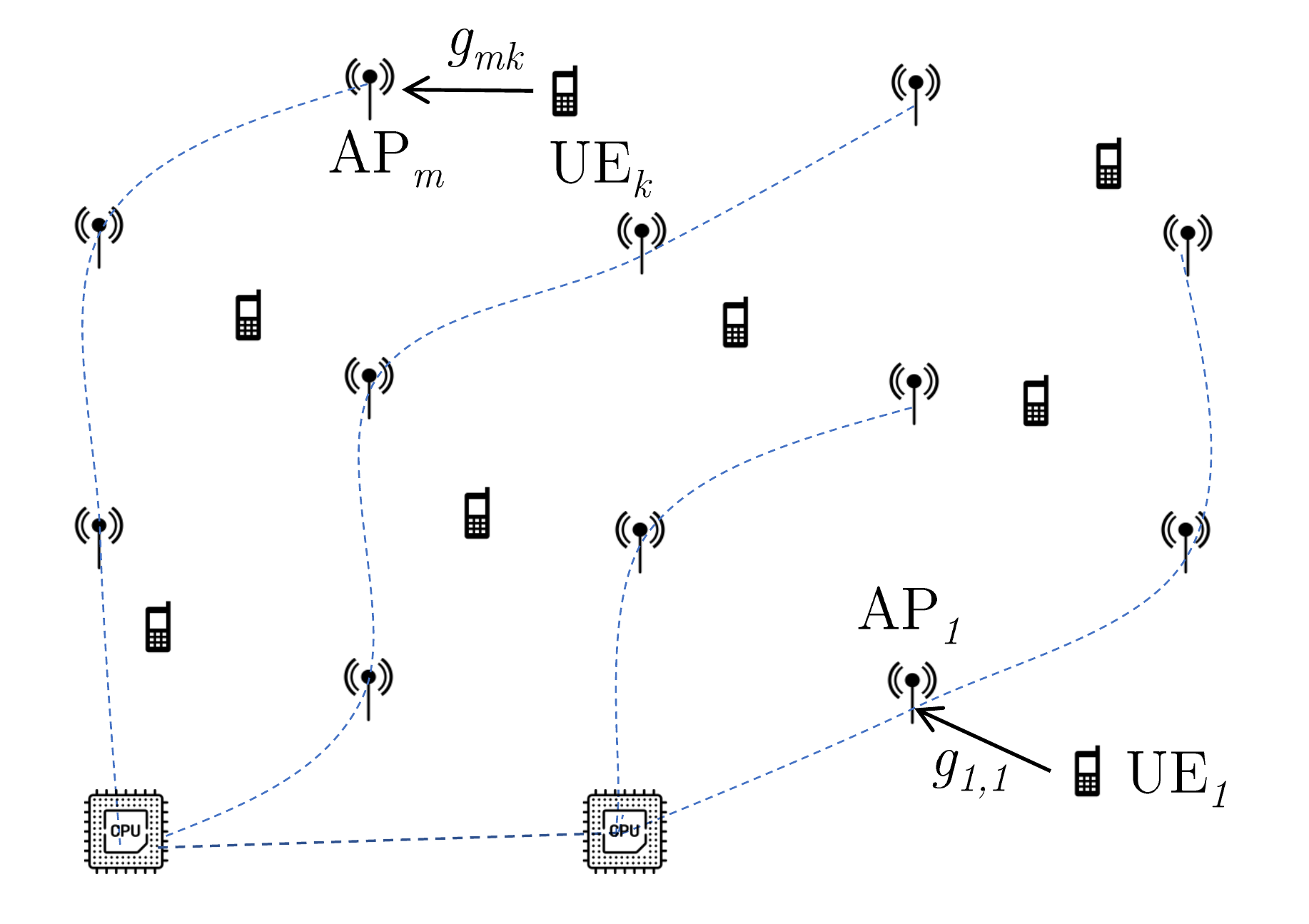}
        \caption{}
        \label{fig:system_model}
    \end{subfigure}
    \begin{subfigure}[b]{\linewidth}
         \centering
        \includegraphics[width=0.65\textwidth]{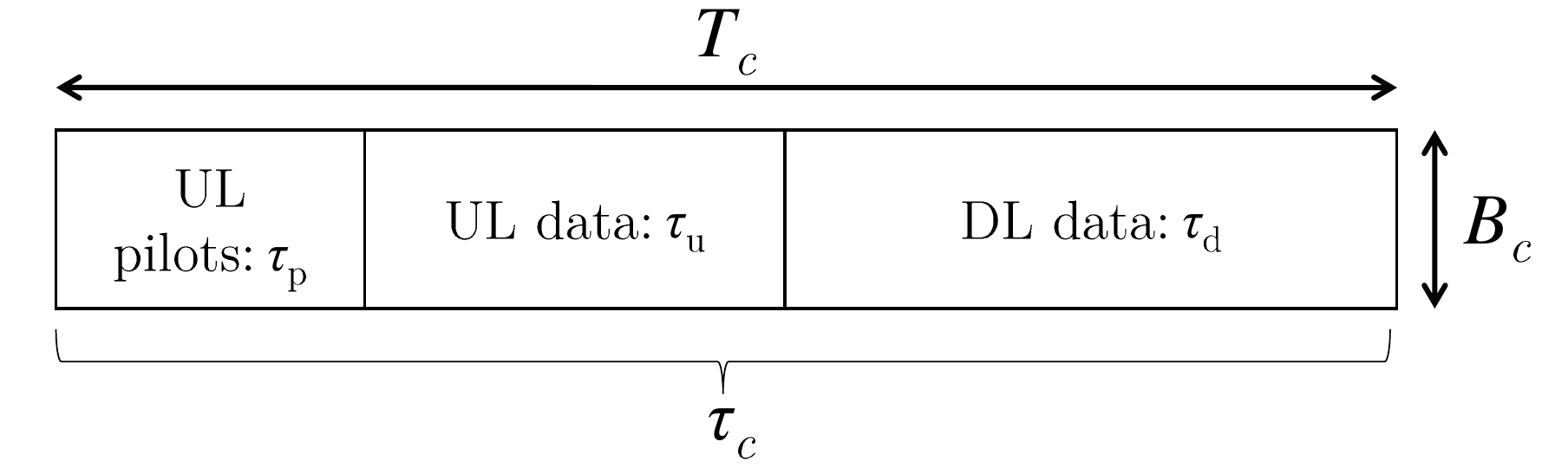}
        \caption{}
        \label{fig:fading_block}
    \end{subfigure}
    \caption{(a) A CF mMIMO system, where $M$ distributed APs jointly serve $K$ UEs ($K\ll M$). (b) A coherence block with UL pilot training, and UL and DL data transmission phases.}
\end{figure}
We consider a typical \gls{cf} \gls{mmimo} system, where $K$ single-antenna \glspl{ue} are jointly served by $M$ geographically distributed \glspl{ap} each equipped with $L$ antennas ($K \ll M$), as shown in~\Cref{fig:system_model}.
The \glspl{ap} are connected to one or multiple \glspl{cpu} by unlimited and error-free fronthaul channels.
The \glspl{cpu} are connected through very fast optical fibers and share the information, and every time one arbitrary \gls{cpu} is responsible for the pilot assignment.
So, our approach is fully centralized, and the study of distributed pilot assignments is left for future work.

The channel coefficient $\textbf{g}_{mk}\in\mathbb{C}^{L\times 1}$ between the $m$-th \gls{ap} and the $k$-th \gls{ue} is given as
\begin{equation}\label{eq:channel_coefficient}
\textbf{g}_{mk} = \beta_{mk}^{1/2}\textbf{h}_{mk},
\end{equation}
where $\{\beta_{mk}\}$ represent the \gls{lsf} coefficients, i.e., pathloss and shadowing, and  $\{\textbf{h}_{mk}\}$ indicate the small-scale fading coefficients which are assumed to be independent and identically distributed (i.i.d.) normal random variables $\mathcal{CN}(0,\textbf{I}_L)$.
We adopt the block fading model, shown in \Cref{fig:fading_block}, where time-frequency resources are divided into coherence intervals of $\tau_c$ channel uses in which the channel can be approximately considered as static.
The duration of the intervals is defined based on propagation environment, \glspl{ue} mobility, and carrier frequency.

Each interval is further divided into three sub-intervals such that: $\tau_c = \tau_p + \tau_u + \tau_d$, where $\tau_p$ is used for \gls{ul} pilot training, and $\tau_u$ and $\tau_d$ are used for \gls{ul} and \gls{dl} data transmission, respectively.
We assume that $\textbf{h}_{mk}$ stays constant during a coherence interval and is independent in different coherence intervals.
We eliminate the \gls{dl} pilot training phase by assuming channel reciprocity, i.e., the same channel coefficients for the \gls{ul} and \gls{dl} transmissions.

\subsection{Uplink Pilot Training}
We assume that there are only $\tau_p$ mutually orthogonal pilot sequences with length $\tau_p$ each represented as a column $\pmb{\phi} \in \mathbb{C}^{\tau_p \times 1}$ of a matrix $\mathbf{\Phi}$, for which we have $\Vert\pmb{\phi}_{p_k}^{H}\pmb{\phi}_{p_{k'}}\Vert = 1$ if $p_k=p_{k'}$, and $\Vert\pmb{\phi}_{p_{k}}^{H}\pmb{\phi}_{p_{k'}}\Vert = 0$, otherwise, and $p_k\in\{1,\ldots,\tau_p\}$ indicates the index of the pilot assigned to the $k$th \gls{ue}.

In the \gls{ul} pilot training phase, all \glspl{ue} simultaneously transmit their pilots.
The $m$-th \gls{ap} receives

\begin{equation}
{\textbf{Y}}_m^\mathrm{p}=\sqrt{\tau_{p} \rho^{\mathrm{p}}} \sum_{k=1}^{K} \textbf{g}_{m k} \boldsymbol{\pmb{\phi}}_{p_{k}}^{H}+\textbf{W}_m^\mathrm{p}, 
\end{equation}
where $\rho^\mathrm{p}$ is the normalized \gls{snr} of a pilot sequence with respect to the noise power, and $\textbf{W}_m^\mathrm{p} $ is the $L\times \tau_c$ additive noise matrix with elements following i.i.d. $\sim\mathcal{CN}(0, 1)$ random variables. 

As shown in~\cite{ngo2017cell, mai2018pilot}, the effective channel coefficients between \gls{ue} $k$ and \gls{ap} $m$ can be estimated employing the \gls{mmse} estimator after projecting $\textbf{Y}_m^\mathrm{p}$ onto $\pmb{\phi}_k^H$ as

\begin{equation}
\hat{\textbf{g}}_{m k}\!=\mathbb{E}\left\{ \textbf{g}_{mk}\check{\textbf{y}}^{\mathrm{p}H}_{mk}\right\}\!\Big(\mathbb{E}\left\{\check{\textbf{y}}_{mk}^{\mathrm{p}}\check{\textbf{y}}^{\mathrm{p}H}_{mk}\right\}\Big)^{-1}\check{\textbf{y}}_{mk}^{\mathrm{p}} = c_{mk}\check{\textbf{y}}_{mk}^{\mathrm{p}},
\end{equation}
where $\check{\textbf{y}}_{mk}^{\mathrm{p}} \triangleq \textbf{Y}^{\mathrm{p}}_m\pmb{\phi}_{p_k}$, and 
\begin{equation}
c_{m k} \triangleq \frac{\sqrt{\tau_{p} \rho^{\mathrm{p}}} \beta_{m k}}{\tau_{p} \rho^{\mathrm{p}} \sum_{k^{\prime}=1}^{K} \beta_{m k^{\prime}}\left|\pmb{\phi}_{p_{k}}^{H} \pmb{\phi}_{p_{k^{\prime}}}\right|^{2}+1}.
\end{equation}

The $l$-th component's mean-square of the estimated channel vector $\hat{\textbf{g}}_{mk}$ can be calculated as
\begin{equation}
\gamma_{m k} \triangleq \mathbb{E}\left\{\left|\left[\hat{\textbf{g}}_{m k}\right]_l\right|^{2}\right\}=\sqrt{\tau_{p} \rho^{\mathrm{p}}} \beta_{m k} c_{m k}.
\end{equation}

\subsection{Uplink Data Transmission}
In \gls{cf} \gls{mmimo}, all \glspl{ap} and \glspl{ue} share the same time-frequency resources for data transmission.
In the \gls{ul}, all \glspl{ue} simultaneously transmit their data to the \glspl{ap}.
The \gls{ap} $m$ receives
\begin{equation}
{\textbf{y}}_m^\mathrm{u}=\sqrt{\rho^\mathrm{u}} \sum_{k=1}^{K} \textbf{g}_{m k} \sqrt{\eta_k}q_k+\textbf{w}_m^\mathrm{u}, 
\end{equation}
where $q_k$ is the signal transmitted by \gls{ue} $k$ with power $\mathbb{E}\left\{\left|q_k\right|^{2}\right\}=1$, $\eta_k\in\left[0,1\right]$ shows the power control coefficient, $\rho^{\mathrm{u}}$ indicates the normalized \gls{ul} \gls{snr} and $w_m^{\mathrm{u}} \sim\mathcal{CN}(0, 1)$ is the additive noise at the receiver.

The \gls{mr} combining scheme can be applied to decode the desired signal for a certain \gls{ue} $k$.
\gls{ap} $m$ sends $\hat{\textbf{g}}^*_{mk}\textbf{y}^{\mathrm{u}}_m$ to the \gls{cpu} for data detection.
The \gls{cpu} combines all the received signals for \gls{ue} $k$ as
\begin{equation}
    r^{\mathrm{u}}_k = \sum_{m=1}^M\sum_{l=1}^L\left[\hat{\textbf{g}}_{mk}\right]^*_l\left[\textbf{y}^{\mathrm{u}}_m\right]_l.
\end{equation}

Following the same procedure as in \cite{ngo2017cell}, the signal then can be decomposed at the \gls{cpu} as
\begin{equation}
\begin{aligned}
r_k^\mathrm{u}=& \underbrace{\sqrt{\rho^{\mathrm{u}} \eta_{k}} q_{k} \mathbb{E}\left\{\sum_{m=1}^{M}\sum_{l=1}^L \left[\hat{\textbf{g}}_{m k}\right]^*_l \left[\textbf{g}_{m k}\right]_l\right\}}_{\texttt{DS}_{k}}\\
&+\underbrace{\sqrt{\rho^{\mathrm{u}} \eta_{k}} q_{k}\left(\sum_{m=1}^{M}\sum_{l=1}^L \left[\hat{\textbf{g}}_{m k}\right]^*_l \left[\textbf{g}_{m k}\right]_l-\mathbb{E}\left\{\sum_{m=1}^{M}\sum_{l=1}^L \left[\hat{\textbf{g}}_{m k}\right]^*_l \left[\textbf{g}_{m k}\right]_l\right\}\right)}_{\texttt{BU}_{k}} \\
&+\underbrace{\sqrt{\rho^{\mathrm{u}}} \sum_{m=1}^{M} \sum_{k^{\prime} \neq k}^{K}\sum_{l=1}^L \sqrt{\eta_{k^{\prime}}} \left[\hat{\textbf{g}}_{m k}\right]^*_l \left[\textbf{g}_{m k^{\prime}}\right]_l q_{k^{\prime}}}_{\texttt{CPI}_{k}}+\sum_{m=1}^M\sum_{l=1}^L\left[\hat{\textbf{g}}_{m k}\right]^*_l \left[\textbf{w}_m^\mathrm{u}\right]_l,
\end{aligned}\label{eq:7}
\end{equation}
where $\texttt{DS}_k$, $\texttt{BU}_k$ and $\texttt{CPI}_k$ denoted the desired signal (DS), beamforming uncertainty (BU) and co-pilot interference (CPI), respectively.

The \gls{ul} achievable rate for \gls{ue} $k$ can be calculated as
\begin{equation}
    \begin{aligned}\label{eq:ul_sinr}
    \operatorname{R}_k^\mathrm{u}=\log_2\left(1+\frac{L^2\rho^{\mathrm{u}} \eta_{k}\left(\sum_{m=1}^M \gamma_{m k}\right)^{2}}{L^2\rho^{\mathrm{u}} \sum_{k^{\prime} \neq k}^{K} \eta_{k^{\prime}}\left(\sum_{m=1}^{M} \gamma_{m k} \frac{\beta_{m k^{\prime}}}{\beta_{m k}}\right)^2\left|\pmb{\phi}_{p_{k}}^{H} \pmb{\phi}_{p_{k^{\prime}}}\right|+L\rho^{\mathrm{u}} \sum_{k^{\prime}=1}^{K} \eta_{k^{\prime}} \sum_{m=1}^{M} \gamma_{m k} \beta_{m k^{\prime}}+L\sum_{m=1}^{M} \gamma_{m k}}\right).
\end{aligned}
\end{equation}

\subsection{Downlink Data Transmission}
In \gls{dl}, \glspl{ap} receive encoded data from their \glspl{cpu} and carry out the transmit precoding, based on the local \gls{csi}. 
The $k$th \gls{ue} receives signal
\begin{equation}
r_k^\mathrm{d} = \sqrt{\rho^\mathrm{d}}\sum_{m=1}^M\sum_{k'=1}^K \eta_{mk'}^{1/2}\textbf{g}^T_{mk}\hat{\textbf{g}}^*_{mk'} q_{k'}+ w_k^\mathrm{d},
\end{equation}
where $\rho^{\mathrm{d}}$ is the normalized \gls{ul} \gls{snr}, and $w_k^{\mathrm{d}} \sim\mathcal{CN}(0, 1)$ is the additive noise at the $k$th \gls{ue}.
Then $q_k$ will be detected from $r_k^{\mathrm{d}}$.

Employing a similar methodology as in the \gls{ul}, the achievable \gls{dl} rate for the $k$th \gls{ue} can be derived from

\begin{equation}
\begin{aligned}\label{eq:dl_sinr}
\operatorname{R}_k^{\mathrm{d}}=\log _{2}\left(1+\frac{L^2\rho^{\mathrm{d}}\left(\sum_{m=1}^{M} \eta_{m k}^{1 / 2} \gamma_{m k}\right)^{2}}{\rho^{\mathrm{d}} \sum_{k^{\prime} \neq k}^{K}\left(\sum_{m=1}^{M} \eta_{m k^{\prime}}^{1 / 2} \gamma_{m k^{\prime}} \frac{\beta_{m k}}{\beta_{m k^{\prime}}}\right)^{2}\left|\pmb{\phi}_{k^{\prime}}^{H} \pmb{\phi}_{k}\right|^{2}+L\rho^{\mathrm{d}} \sum_{k^{\prime}=1}^{K} \sum_{m=1}^{M} \eta_{m k^{\prime}} \gamma_{m k^{\prime}} \beta_{m k}+1}\right).
\end{aligned}
\end{equation}
\section{Pilot Assignment in CF mMIMO}\label{sec:problem_formulation}
\subsection{Problem formulation}
The goal of the \gls{ul} pilot training phase is to increase the number of effectively estimated channels or to improve the quality of the \gls{ul} channel estimation, which can be interpreted as a \gls{ul} rate maximization problem.
So, an efficient pilot reuse scheme should assign pilots to \glspl{ue} so that the sum of the \gls{ul} rates is maximized.
Mathematically the pilot reuse problem in \gls{cf} \gls{mmimo} can be formulated as

\begin{maxi!}|s|[1]
    {\textbf{p}}
    {\sum_{k=1}^K \operatorname{R}_k^\mathrm{u}}{}{}
    \addConstraint{\textbf{p}}{=[p_1,\ldots,p_K]^{T}}{}
    \addConstraint{\pmb{\phi}_{p_k}}{=\mathrm{col}_{p_k}(\pmb{\Phi})}{}
    \addConstraint{p_k}{\in\{1,\ldots,\tau_p\},}{}
\end{maxi!}
where the $\textbf{p}$ is the pilot assignment vector and $\mathrm{col}_i(\pmb{\Phi})$ indicates the $i$th column of matrix $\pmb{\Phi}$.

The co-pilot interference originated from reusing the same pilot for similar \glspl{ue}, i.e., geographical closeness, common serving \glspl{ap}, and similar channel coefficients.
So, an intelligent pilot assignment scheme should consider the similarity among the \glspl{ue} and reuse the same pilot $\pmb{\phi}$ in \glspl{ue} that have higher dissimilarities, i.e., geographically far apart or with fewer common serving \glspl{ap}.
We hence formulate the pilot assignment in \gls{cf} \gls{mmimo} as a diverse clustering problem, where we form $\tau_p$ (number of available orthogonal pilots) clusters in such a way that \glspl{ue} belonging to the same cluster have a high ``dissimilarity''  or ``diversity''.
In the following subsection, we formulate and discuss the problem.

\subsection{Diverse Clustering Problem} \label{sec:dcc}
The \gls{dcp} consists of the assigning of a set of $K$ elements, i.e., \glspl{ue}, into $\tau_p$ mutually disjoint subsets or clusters, i.e., pilots, while the diversity among the elements in each subset, i.e.,  intra-cluster diversity, and inter-cluster similarity is maximized.
The inter-cluster diversity is calculated as the sum of the individual distances between each pair of elements in clusters, where the concept of distance depends on the specific application context.
The objective is to maximize the overall diversity, i.e., the sum of the diversity of all subsets.
From the graph theory perspective, \gls{dcp} can be considered as partitioning the vertices of a complete weighted undirected graph into $\tau_p$ subgraphs in such a way that the total weight of the subgraphs is maximized while applying optional constraints on the number of nodes in each subgraph.
\begin{figure}
     \centering
     \begin{subfigure}[b]{0.32\textwidth}
         \centering
         \includegraphics[width=\textwidth]{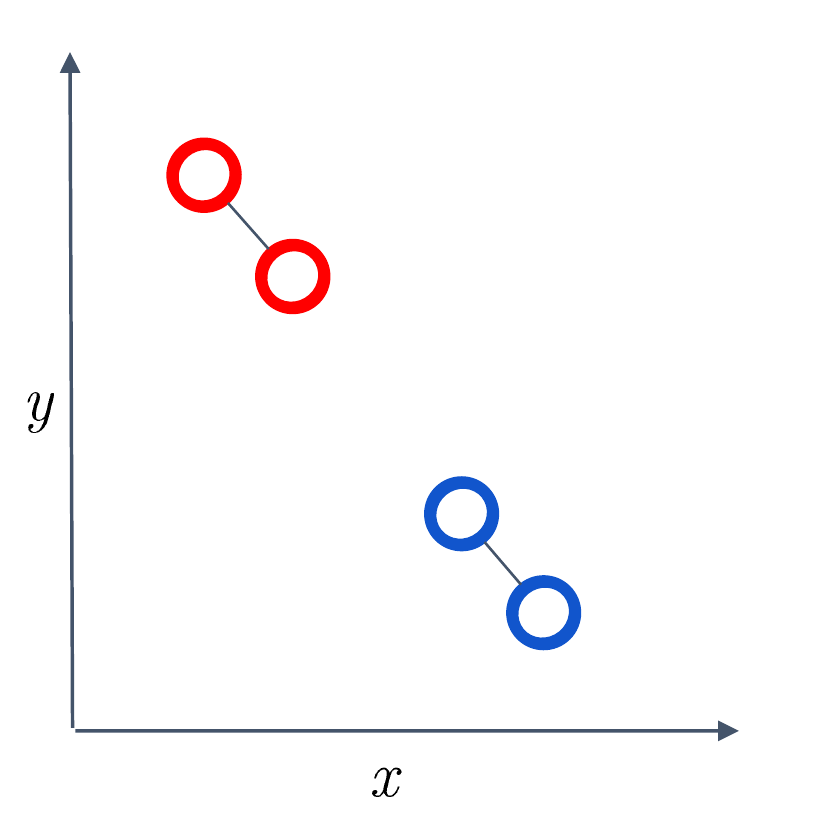}
         \caption{}\label{fig:dcp-a}
     \end{subfigure}
     \hfill
     \begin{subfigure}[b]{0.32\textwidth}
         \centering
         \includegraphics[width=\textwidth]{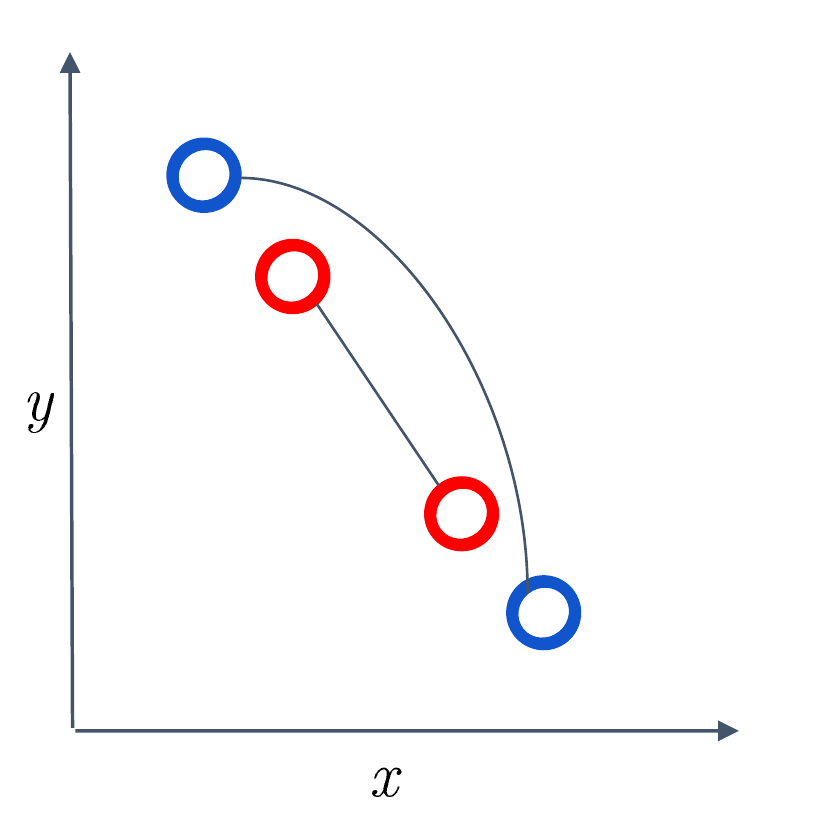}
         \caption{}\label{fig:dcp-b}
     \end{subfigure}
     \centering
     \begin{subfigure}[b]{0.32\textwidth}
         \centering
         \includegraphics[width=\textwidth]{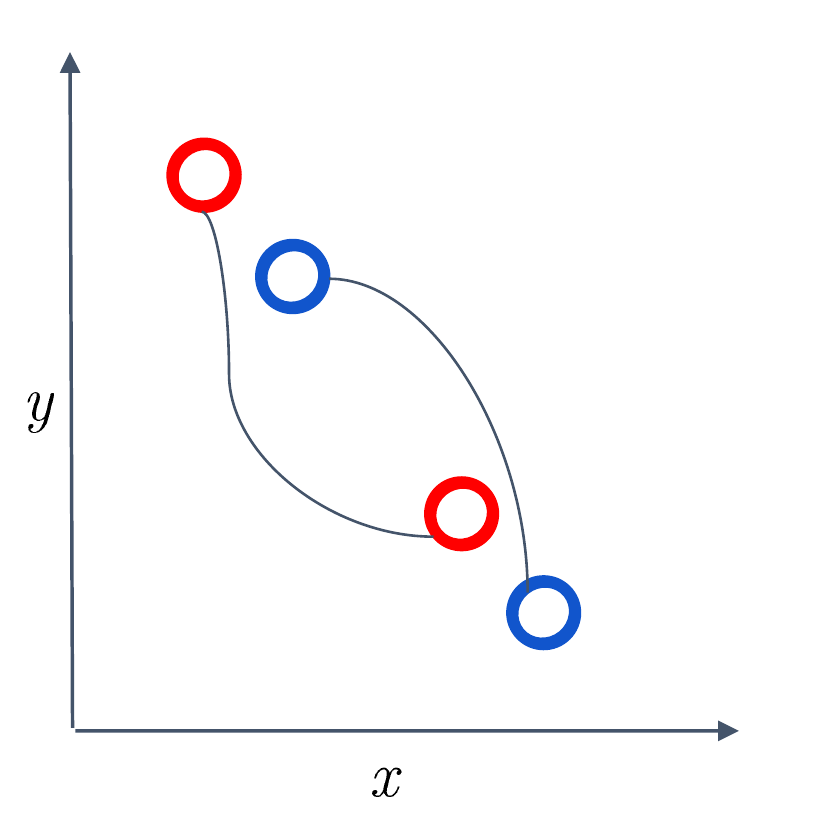}
         \caption{}\label{fig:dcp-c}
     \end{subfigure}
     \hfill
        \caption{Illustrated example and comparison of the conventional clustering (a), and feasible (b) and optimal (c) solutions for diverse clustering, with four data points and two clusters.}
        \label{fig:dcp}
\end{figure}

An illustration of \gls{dcp} and of the conventional clustering problem is presented in \Cref{fig:dcp} for a small configuration with four data points and two clusters.
\Cref{fig:dcp-a} shows a conventional clustering problem, where clusters are formed in such a way to minimize inter-cluster similarity (intra-cluster diversity).

In contrast, \Cref{fig:dcp}b-c show two \glspl{dcp}, where data points with higher dissimilarities are joining the same cluster.
\Cref{fig:dcp-b} represents a feasible solution for \gls{dcp}, in which the diversity score of one cluster (blue data points) is relatively higher than the other, while in the optimal solution, all clusters should have relatively similar diversity score, as it is the case in \Cref{fig:dcp-c}.

In general \gls{dcp} can be considered as a capacitated clustering~\cite{zhou2019heuristic, brimberg2019solving} or maximally diverse grouping problem~\cite{schulz2021balanced, brimberg2015solving} and then formulated as a quadratic integer program as
\begin{maxi!}|s|[1]
    {\textbf{X}}
    {\sum_{p=1}^{\tau_p} \sum_{k=1}^{K-1}\sum_{k\prime=k+1}^{K} x_{k,p}x_{k',p}d_{k,k'}}{}{}\label{eq:dcc}
    \addConstraint{\sum_{p=1}^{\tau_p}x_{k,p}}{=1, \quad \label{eq:dcc_con1}}{k=1,\ldots,K}
    \addConstraint{L_k \leq \sum_{k=1}^K x_{k,p}}{\leq U_k,\quad \label{eq:dcc_con2}}{p=1,\ldots,\tau_p}
    \addConstraint{x_{k,p}}{\in\{0,1\}, \quad \label{eq:dcc_con3}}{ k=1,\ldots,K,  p=1,\ldots,\tau_p,}
\end{maxi!}
where $\textbf{X}\in\{0,1\}^{K\times \tau_p}$ is a binary association matrix, and $x_{k,p}=1$ if element (\gls{ue}) $k$ belongs to cluster (pilot) $p$, and $x_{k,p}=0$ otherwise. $d_{k, k'}$ is the diversity measure between $k$ and $k'$ elements, and $L_b$ and $U_b$ show the minimum and maximum size of each set, respectively.
The constraint \eqref{eq:dcc_con1} guarantees that each element is assigned to only one cluster, and \eqref{eq:dcc_con2} forces the size of the clusters to be in the specified range.
The diversity measure can be a predefined distance function, i.e., Euclidean distance and cosine similarity, or can be defined as a parameterized kernel and then learned by, e.g., neural networks.

This formulation favors forming fewer large-size clusters against many small-size clusters.
Considering the full interference among the nodes in each cluster (orthogonal pilot), in the pilot reuse problem, fewer large-size clusters increase the co-pilot interference among the co-cluster \glspl{ue}.
So we multiply a regularization term by the objective of the optimization problem to penalize the large-size clusters by dividing their score by the size of clusters.
This will avoid wasteful growth of the size of some clusters.
Adding a new node to a $N$-size cluster is interpreted as interference with all $N$ nodes and penalized.
Hence, the new formulation will be

\begin{maxi}|s|[1]
    {\textbf{X}}
    {\sum_{p=1}^{\tau_p} \frac{1}{|\mathcal{C}_p|}\sum_{k=1}^{K-1}\sum_{k\prime=k+1}^{K} x_{k,p}x_{k',p}d_{k,k'}}{}{}
    \addConstraint{\eqref{eq:dcc_con1} - \eqref{eq:dcc_con3}}
\end{maxi}
where $\mathcal{C}_p$ is the cluster set $p$, and $|\cdot|$ shows the size of a set.

\gls{dcp} is a combinatorial optimization problem and is proved to be NP-hard~\cite{feo1990class}.
Typically finding the exact solution for these problems is not computationally possible, at least when $K$ is large.

\gls{dcp} has already been investigated in the literature under different names, such as maximally diverse grouping problem ~\cite{schulz2021balanced} and anticlustering~\cite{papenberg2021using, brusco2020combining}.
Several algorithms have already been proposed to solve \gls{dcp}, including tabu search with strategic oscillation  \cite{gallego2013tabu}, genetic algorithm~\cite{fan2011erratum,singh2019new}, artificial bee colony~\cite{rodriguez2013artificial}, variable neighborhood search~\cite{brimberg2015solving}.
In this paper, we propose an iterative maxima search method, adopted from~\cite{lai2016iterated}, for pilot reuse in~\gls{cf} \gls{mmimo} based on \gls{dcp} problem. 

\section{Iterative Maxima Search for Pilot Reuse}\label{sec:iterative_optima_search}
The proposed approach employs an iterative maxima search procedure that integrates a local neighborhood search procedure with a weak perturbation operator to improve the intensity or quality of solutions and a robust perturbation operator to improve the diversity of the solutions by moving the search to a distant region to avoid local optimum solutions.
Before going to the details of the proposed scheme, some concepts need to be defined.

\subsubsection*{Definition of neighborhoods}
We define two different types of neighborhood: \textit{OneMove} ($N_1$) neighborhood and \textit{SwapMove} ($N_2$) neighborhood.
Given the pilot assignment vector $\textbf{p}$\footnote{The pilot assignment vector \textbf{p} is equivalent to the association matrix \textbf{X} in \eqref{eq:dcc}, and in fact, \textbf{X} is the one-hot encoding version of \textbf{p}.}, $N_1$ returns all possible solutions obtained by changing the assigned pilot of a single \gls{ue} (\textit{OneMove}) in such a way that the pilot reuse capacity constraints are satisfied, while $N_2$ returns the possible solutions obtained by exchanging the pilots of a pair of \glspl{ue} (\textit{SwapMove}).

\subsubsection*{\textbf{M} matrix} 
To improve the computational efficiency of the local search procedure, we employ a $K\times \tau_p$ matrix $\textbf{M}$, where $m_{k,p}\in\textbf{M}$ shows the sum of diversity between \gls{ue} $k$ and all \glspl{ue} with pilot index $p$, and having the pilot assignment vector $\textbf{p}$, calculates as $m_{k,p}=\sum_{\substack{k,k'=1,\\ p_{k'}=p}}^Kd_{k,k'}$. 
Calculation of this matrix can be done in order of $\mathcal{O}(K^2)$.

\subsubsection*{Definition of a solution}
The tuple $<\textbf{p}, \mathbf{c}, \mathbf{s}>$ refers to a solution in search space, where $\mathbf{c}$ is a vector that saves the diversity index of each cluster, and $\mathbf{s}$ is a vector that stores the size of each cluster for a given solution.
Basically, for each solution in the search space, we save and update the tuple where the two last elements are used to speed up the search procedure, as we will discuss later.

The overall procedure is presented in~\Cref{alg:global_search}.
The algorithm starts by generating $I_s$ random initial feasible solutions (\Cref{alg:initial_feasible_solution}), followed by a local neighborhood search procedure (\Cref{alg:local_search}).
The best solution among the initial solutions is then saved for later use.
It then repeats a maxima search procedure (lines 9-20) followed by a robust perturbation procedure until a certain time budget is exceeded.
This iterative maxima search procedure is composed of a weak perturbation procedure (\Cref{alg:weak_perturbation}) followed by a local search procedure (\Cref{alg:local_search}), which will be discussed in the following subsections.
In each iteration, after employing the weak perturbation and local search procedures, the quality or fitness of the current solution $\textbf{p}$ ($f(\textbf{p})$) is compared and used to replace the incumbent solution $\textbf{p}^*$ in case of improvement (lines 14-19).
The fitness of a solution is calculated as:
\begin{equation}\label{eq:fitness}
    f(\textbf{p}) = \sum_{p=1}^{\tau_p}\frac{1}{|\mathcal{C}_p|}\sum_{\substack{p_k=p_{k'}, \\k<k'}}d_{kk'},
\end{equation}
where, $d_{kk'}$ represents the diversity among $k$ and $k'$ \glspl{ue}.
This fitness function basically is a weighted average of the diversity score of different clusters, where the weight is the inverse of the cluster size.

\begin{algorithm}[t]
    \caption{Iterative Maxima Search for Pilot Reuse} 
    \label{alg:global_search}
    \textbf{Input:} $I_s$, $t_{max}$, $\alpha$
     \;\\
    \textbf{Output:} Pilot assignment vector $\textbf{p}^*$
    \begin{algorithmic}[1]
        \State Initialize empty set $\mathcal{P}_0$
        \For{$i=1\ \textbf{to}\ I_s$}
        \State Generate initial solution $\textbf{p}^0$ employing \Cref{alg:initial_feasible_solution}.
        \State Update $\textbf{p}^0$ employing the local search in \Cref{alg:local_search}.
        \State Add $\textbf{p}^0$ to $\mathcal{P}_0$.
        \EndFor
        \State Get the best solution from $\mathcal{P}_0$ and save it in $\textbf{p}$
        \State $\textbf{p}^* \gets \textbf{p}$
        \While{$Time() \leq t_{max}$}
            \State Set $ctr \gets 0$
            \While{$ctr<\alpha$}
            \State Apply weak perturbation operator in \Cref{alg:weak_perturbation} on $\textbf{p}$.
            \State Update $\textbf{p}$ employing local search \Cref{alg:local_search}.
            \If{$f(\textbf{p}) > f(\textbf{p}^*)$}
                \State $\textbf{p}^* \gets \textbf{p}$
                \State Reset $ctr \gets 0$
            \Else
                \State Increase $ctr \gets ctr + 1$
            \EndIf
            \EndWhile
            \State Update $\textbf{p}$ employing the robust perturbation in \Cref{alg:strong_perturbation}.
        \EndWhile
        \State Return $\textbf{p}^*$
    \end{algorithmic}
\end{algorithm}

\subsection{Initial Feasible Solution}
The Initial feasible solution procedure is presented in \Cref{alg:initial_feasible_solution}.
The procedure starts by randomly assigning each pilot to $L_b$ \glspl{ue} and then for the remaining \glspl{ue} assigns a random pilot $p$ while being sure that the number of \glspl{ue} with pilot $p$, $|\mathcal{C}_p|$, is less than an upper bound $U_b$. The complexity of this algorithm is $\mathcal{O}(K+\tau_p)$.

\begin{algorithm}[t]
    \caption{Initial Feasible Solution} 
    \label{alg:initial_feasible_solution}
    \textbf{Input:}
     Set of Pilots $\mathcal{P}$, Set of UEs $\mathcal{K}$, $L_b$, $U_b$\;\\
    \textbf{Output:}
    Initial Pilot Assignment Vector $\textbf{p}$
    \begin{algorithmic}[1]
        \For{$p\in\mathcal{P}$}
            \State Randomly select $L_b$ UEs from $\mathcal{K}$ and name it $\mathcal{K'}$\;
            \State Assign pilot $p$ to UEs in $\mathcal{K'}$ and update $\mathcal{K} = \mathcal{K} \setminus \mathcal{K'}$\;
        \EndFor
        \While{$\mathcal{K}\neq\emptyset$}
            \State Get $k\in\mathcal{K}$ and randomly select pilot $p\in\mathcal{P}$\;
            \If{$|\mathcal{C}|_p<U_b$}
            \State Assign pilot $p$ to $k$ and update $\mathcal{K}=\mathcal{K}\setminus \{k\}$\;
            \EndIf
        \EndWhile
        \State Return assignment vector \textbf{p}.
    \end{algorithmic}
\end{algorithm}

\subsection{Local Neighborhood Search}
The local neighborhood search procedure is presented in \Cref{alg:local_search}.
Given the current assignment $\textbf{p}$, the procedure probes $N_1(\textbf{p})$ (lines 3-11) and $N_2(\textbf{p})$ (lines 12-19) neighborhoods and iteratively updates the incumbent solution with the better neighbor solution.
The procedure repeats until the incumbent solution finds the local optimum and can not be further improved.
Given a solution $<\textbf{p}, \mathbf{c}, \mathbf{s}>$, the fitness of a neighbor solution can be easily computed using the defined above matrix $\textbf{M}$.
For $N_1$ neighbors, changing the pilot index of \gls{ue} $k$ from $i$ to $j$ does not change the diversity values of the \glspl{ue}, except those with pilot index $i$ and $j$.
Here the value of a \textit{OneMove} can be determined by
\begin{equation}\label{eq:delfaFOneMove}
    \Delta f = \frac{c_j+m_{kj}}{s_j+1} - \frac{c_j}{s_j} + \frac{c_i-m_{ki}}{s_i-1} - \frac{c_i}{s_i}, 
\end{equation}
where $m_{k,i}$ and $m_{k,j}$ are the entries of matrix $\textbf{M}$ and $c$, and $s$ are the elements of vectors $\textbf{c}$ and $\textbf{s}$, respectively.

Also for $N_2$ neighbors, the value of a \textit{SwapMove} (exchanging the pilot index of \glspl{ue} $k$ and $k'$) is determined by 
\begin{equation}\label{eq:delfaFSwapMove}
    \Delta f = (m_{k,p_{k'}} - m_{k,p_{k}) + (m_{k',p_k}-m_{k',p_{k'}})-2d_{kk'}}.
\end{equation}

 \begin{algorithm}[t]
    \caption{Local Search Procedure} 
    \label{alg:local_search}
    \textbf{Input:}
    Set of Pilots $\mathcal{P}$, Set of UEs $\mathcal{K}$, \textbf{p}, $L_b$, $U_b$\;\\
    \textbf{Output:}
    Local optimum assignment $\textbf{p}^{*}$\;
    \begin{algorithmic}[1]
        \State $\textbf{p}^{*} \gets \textbf{p}$
        \State Initialize $\textbf{M}\in \mathbb{R}^{K\times\tau_p}$\;
        \While{solution improves}
            \For{$k\in\mathcal{K} \land p\in\mathcal{P}$}
                \If{$(p^*_k\neq p)\land (\mathcal{C}_{p^{*}_k}>L_b) \land (\mathcal{C}_p<U_b)$}
                    \State Calculate $\Delta f$ by \eqref{eq:delfaFOneMove}
                    \If{$\Delta > 0$}
                        \State Set $p^*_k \gets p$ and update $\textbf{M}$
                    \EndIf
                \EndIf
            \EndFor
            \For{$k,k'\in\mathcal{K}\land k'>k$}
                \If{$p^*_k \neq p^{*}_{k'}$}
                    \State Calculate $\Delta f$ by \eqref{eq:delfaFSwapMove}
                    \If{$\Delta f > 0$}
                        \State Swap the pilots of $k$ and $k'$ and update $\textbf{M}$
                    \EndIf
                \EndIf
            \EndFor
        \EndWhile
        \State Return $\textbf{p}^{*}$
    \end{algorithmic}
\end{algorithm}

\subsection{Weak and robust perturbation}
The weak and robust perturbation procedures are presented in \Cref{alg:weak_perturbation} and \Cref{alg:strong_perturbation}, respectively.
The weak perturbation operator aims to jump out of the current local optimum within the iterative search procedure by applying some assignment deterioration.
The strength of the weak perturbation is controlled by $\eta_w$, representing the number of random neighbor solutions to be checked by this operator.
For each perturbation step, the best solution among $\eta_{w2}$ randomly selected neighbors is compared, and the incumbent solution is replaced in case of improvement (lines 3-7).
This incumbent solution is used for the next iteration of perturbation.
The large values of $\eta_{w2}$ cause less deterioration of the current sample and can be set to $K$ to adjust the problem size.

\begin{algorithm}[t]
    \caption{Weak Perturbation} 
    \label{alg:weak_perturbation}
    \textbf{Input:} Pilot assignment vector $\textbf{p}$, $\eta_w$, $\eta_w2$\\
    \textbf{Output:} Perturbed assignment $\textbf{p}$
    \begin{algorithmic}[1]
        \For{$i=1\ \textbf{to}\ \eta_w$}
        \State \parbox[t]{\dimexpr\linewidth-2\leftmargin-\labelsep-\labelwidth}{Randomly select a neighbor solution $\textbf{p}'\in N_1(\textbf{p})\cup N_2(\textbf{p})$\strut}
            \For{$j=1\ \textbf{to}\ \eta_{w2}$}
            \State \parbox[t]{\dimexpr\linewidth-2\leftmargin-\labelsep-\labelwidth}{Randomly select a neighbor solution $\textbf{p}''\in N_1(\textbf{p})\cup N_2(\textbf{p})$\strut}
                \If{$f(\textbf{p}'')>f(\textbf{p'})$}
                    \State $\textbf{p}' \gets \textbf{p}''$
                \EndIf
            \EndFor
            \State $\textbf{p}\gets \textbf{p}'$
        \EndFor
        \State Return assignment vector \textbf{p}.
    \end{algorithmic}
\end{algorithm}

The weak perturbation helps the search procedure discover the neighborhood of the current area better, while it is still possible that the search is trapped in a deep local optimum that weak perturbation can not jump out of.
The robust perturbation procedure consequently performs $\eta_s$ moves regardless of their values.
$\eta_s$ controls the strength of the robust perturbation and is empirically set to $\eta_{s}=\theta \times \frac{K}{\tau_p}$, from \cite{lai2016iterated}, where $\theta$ is chosen from $\{1, 1.5\}$.

\begin{algorithm}[t]
    \caption{Robust Perturbation} 
    \label{alg:strong_perturbation}
    \textbf{Input:} Pilot assignment vector $\textbf{p}$, $\eta_s$\\
    \textbf{Output:} Perturbed assignment $\textbf{p}$
    \begin{algorithmic}[1]
        \For{$i=1\ \textbf{to}\ \eta_w$}
            \State \parbox[t]{\dimexpr\linewidth-2\leftmargin-\labelsep-\labelwidth}{Randomly select a neighbor solutions $\textbf{p}'\in N_1(\textbf{p})\cup N_2(\textbf{p})$\strut}
            \State $\textbf{p}\gets \textbf{p}'$
            
        \EndFor
        \State Return assignment vector \textbf{p}.
    \end{algorithmic}
\end{algorithm}

\section{Numerical results}\label{sec:result}
\subsection{Simulation setup}
Let us consider $M$ \glspl{ap} with $L$ antennas and $K$ single antenna \glspl{ue} that are independently and uniformly distributed in a $1\times 1$ km\textsuperscript{2} square area.
The wraparound technique is adopted to mitigate the network edge and boundary effects and  to model the network as if operating over an unlimited area.
The large-scale fading coefficient $\beta_{mk}$ in \eqref{eq:channel_coefficient} is calculated by $\beta_{mk} = \mathrm{PL}_{mk}\cdot 10^{\sigma_{sh}z_{mk}}/10$, where $10^{\sigma_{sh}z_{mk}}/10$ represents the shadow fading with standard deviation $\sigma_{sh}$ and $z_{mk}\sim\mathcal{N}(0,1)$ and $\mathrm{PL}_{mk}$ represents the pathloss from \gls{ue} $k$ to \gls{ap} $m$.
In this paper, we use the three-slope path loss model presented in~\cite{ngo2017cell} as
\begin{equation}
\mathrm{PL}_{m k}=\left\{\begin{array}{l}
-L-15 \log _{10}\left(d_{1}\right)-20 \log _{10}\left(d_{0}\right), \text { if } d_{m k} \leq d_{0} \\
-L-15 \log _{10}\left(d_{1}\right)-20 \log _{10}\left(d_{m k}\right) \\
\text { if } d_{0}<d_{m k} \leq d_{1}\\
-L-35 \log _{10}\left(d_{m k}\right), \text { if } d_{m k}>d_{1}

\end{array}\right.
\end{equation}
where $d_{mk}$ indicates the distance between \gls{ap} $m$ and \gls{ue} $k$, $d_0$ and $d_1$ are the distance thresholds, and $L$ is given by
\begin{equation}\label{eq:pathloss}
\begin{aligned}
L \triangleq &\quad 46.3+33.9 \log _{10}(f)-13.82 \log _{10}\left(h_{\mathrm{AP}}\right) \\
&\quad -\left(1.1 \log _{10}(f)-0.7\right) h_{\mathrm{u}}+\left(1.56 \log _{10}(f)-0.8\right),
\end{aligned}
\end{equation}
where $f$ (MHz) is the carrier frequency, and $h_{\mathrm{u}}$ (m) and $h_{\mathrm{AP}}$ (m) indicate the \gls{ue} and \gls{ap} height respectively.

Noise power is calculated by $P_n=Bk_BT_0W$, where $B$ is the bandwidth, $k_B$ denotes the Boltzmann constant, $T_0$ is the noise temperature and $W$ represents the noise figure.
The transmission powers of the uplink pilot and the uplink data and downlink data are set to $\rho^\mathrm{p} = 100$ (mW), $\rho^{\mathrm{u}} = 100$ (mW), and $\rho^{\mathrm{d}} = 200$ (mW), respectively.
The channel estimation overhead has been taken into account in defining the per-user uplink throughput as $T_k^\mathrm{u} = B\frac{1-\tau_p/\tau_c}{2}\log_2(1+\operatorname{SINR}^\mathrm{u}_k)$, where $\tau_c=200$ samples. The $1/2$ in the above equation is due to the co-existence of the uplink and downlink traffic.
We also employed max-min power control~\cite{ngo2017cell} to further improve the sum throughput.

In this paper we consider the Euclidean distance for the diversity measure as $d_{k,k'} = \sqrt{\sum_{i=1}^{|F|} (U_k[i] - U_{k'}[i])^2}$,  where $F$ is the feature set (e.g., geographical coordinates) of the \glspl{ue}.
The definition and analysis of more sophisticated repulsive functions are left for future work. 

\begin{table}[]
    \centering
    \caption{Simulation setup.}
    \begin{tabular}{lll}\hline
        Parameter & Value & Definition \\\hline
        $B$ & 20 MHz & Bandwidth \\
        $h_{\mathrm{u}}$ & 1.65 m & UE height \\
        $h_{\mathrm{AP}}$ & 15 m & AP height \\
        $d_0$, $d_1$ & 10 m, 50 m & Path loss distance thresholds \\
        $k_B$ & $1.381\times 10^{-23}$ (Joule per Kelvin) & Boltzmann constant \\
        $T_0$ & 290 (Kelvin) & Noise temperature \\
        $W$ & 9 &  Noise figure \\
        $\rho^{\mathrm{p}}$ & 100 mW & Pilot transmission power \\
        $\rho^{\mathrm{u}}$ & 100 mW & Uplink transmission power \\
        $\rho^{\mathrm{d}}$ & 200 mW & Downlink transmission power \\
        \hline
    \end{tabular}
    \label{tab:silulation_setup}
\end{table}

\subsection{Result and discussion}

\begin{figure}
     \centering
     \begin{subfigure}[b]{0.53\textwidth}
         \centering
         \includegraphics[width=\textwidth]{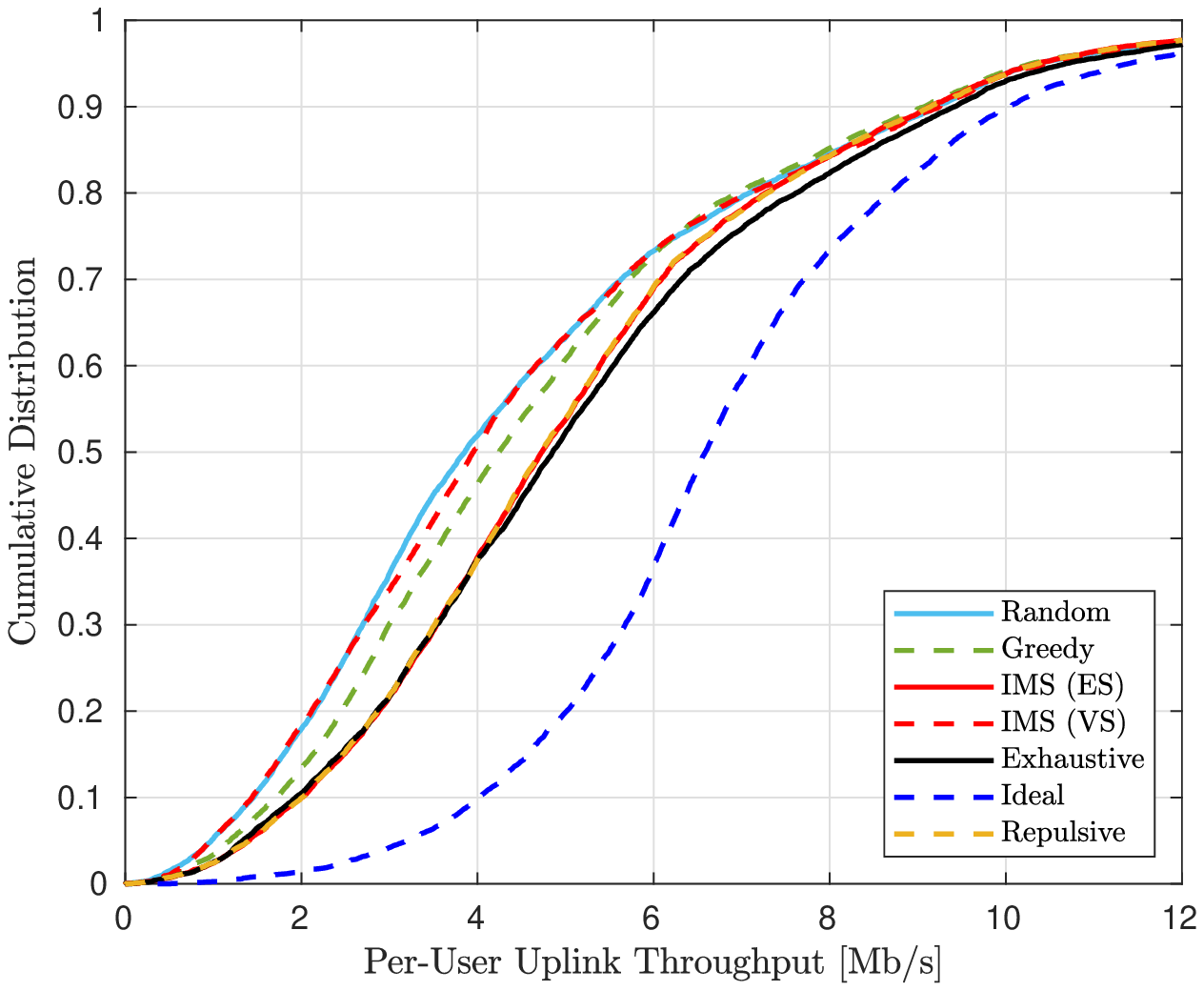}
         \label{fig:cdfUL}
     \end{subfigure}
     \hfill
     \begin{subfigure}[b]{0.53\textwidth}
         \centering
         \includegraphics[width=\textwidth]{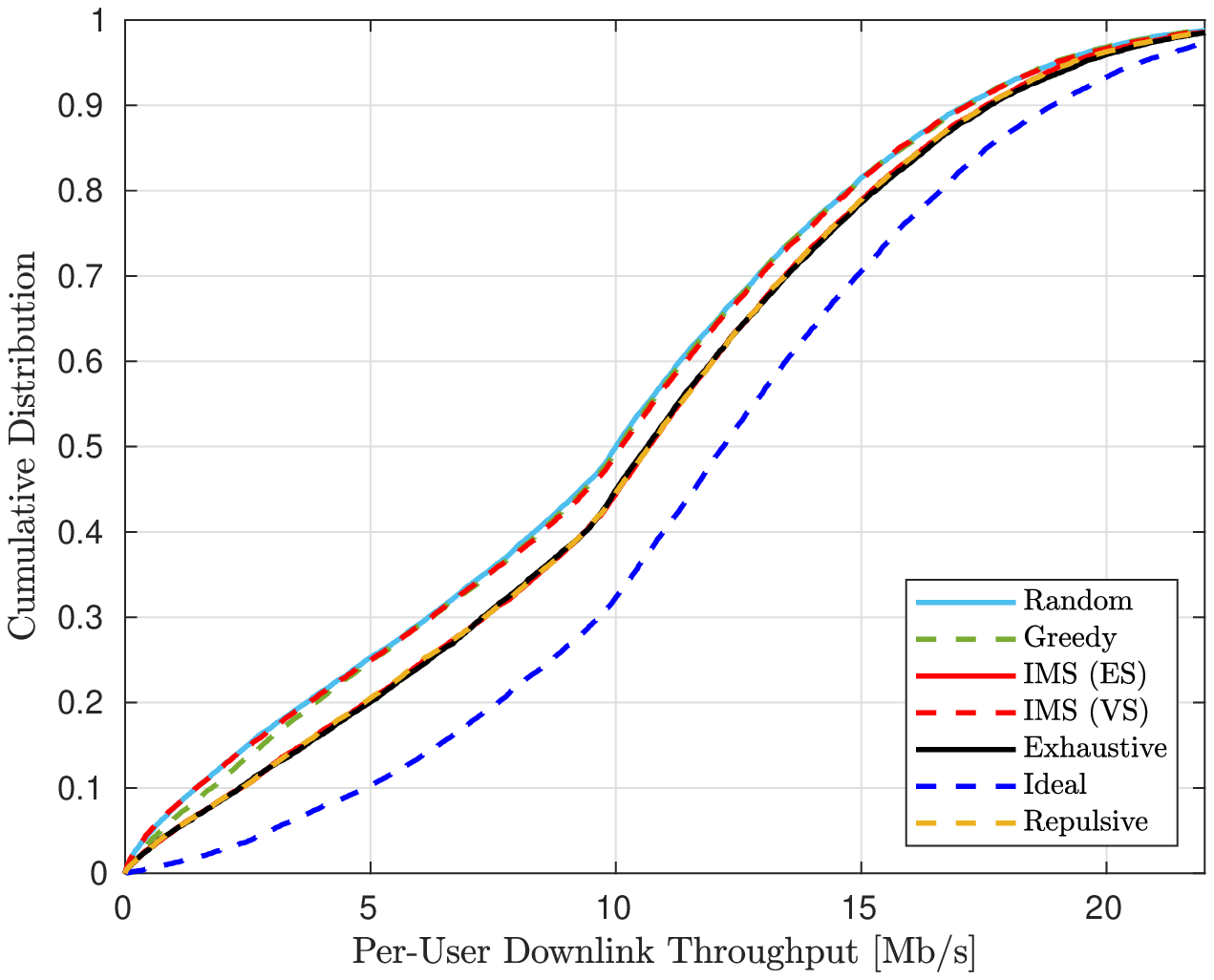}
         \label{fig:cdfDL}
     \end{subfigure}
        \caption{Cumulative distribution of the per-user uplink and downlink throughput for different pilot assignment strategies for a small-scale scenario, $M=50$, $K=12$, $L=1$, and $\tau_p=3$.}
        \label{fig:smallCdf}
\end{figure}
    
This section compares the result of the proposed Iterative Maxima Search (IMS) scheme against different pilot assignment strategies.
In particular, the \textit{random} and \textit{greedy} pilot assignments from \cite{ngo2017cell}, the \textit{repulsive} clustering~\cite{mohebi2022repulsive} and the \textit{Ideal} solution are chosen for evaluation.
The \textit{ideal} solution represents the unreachable upper bound, where there is no pilot contamination (i.e., $\mathrm{CPI}_{k}=0$ in \eqref{eq:7}) and the channels can be perfectly estimated only having one single pilot.
Two different variants of our algorithm are considered: equal size (ES) clusters and variable size (VS) clusters.
The former keeps $L_b=U_b$, while the latter does not have such constraint, and the algorithm can form clusters of any size.

\Cref{fig:smallCdf} shows the per-user throughput \gls{cdf} of different pilot reuse policies for the small-scale scenario for the sake of comparison with the exhaustive search. 
(As the complexity of exhaustive search grows exponentially with the number of \glspl{ue}, calculating its performance for large $M$ is not possible.)
The figure shows that the proposed method outperforms other approaches both in \gls{ul} and \gls{dl} and works almost as well as the optimal pilot reuse obtained by exhaustive search, but with far less complexity.

\begin{figure}
     \centering
     \begin{subfigure}[b]{0.53\textwidth}
         \centering
         \includegraphics[width=\textwidth]{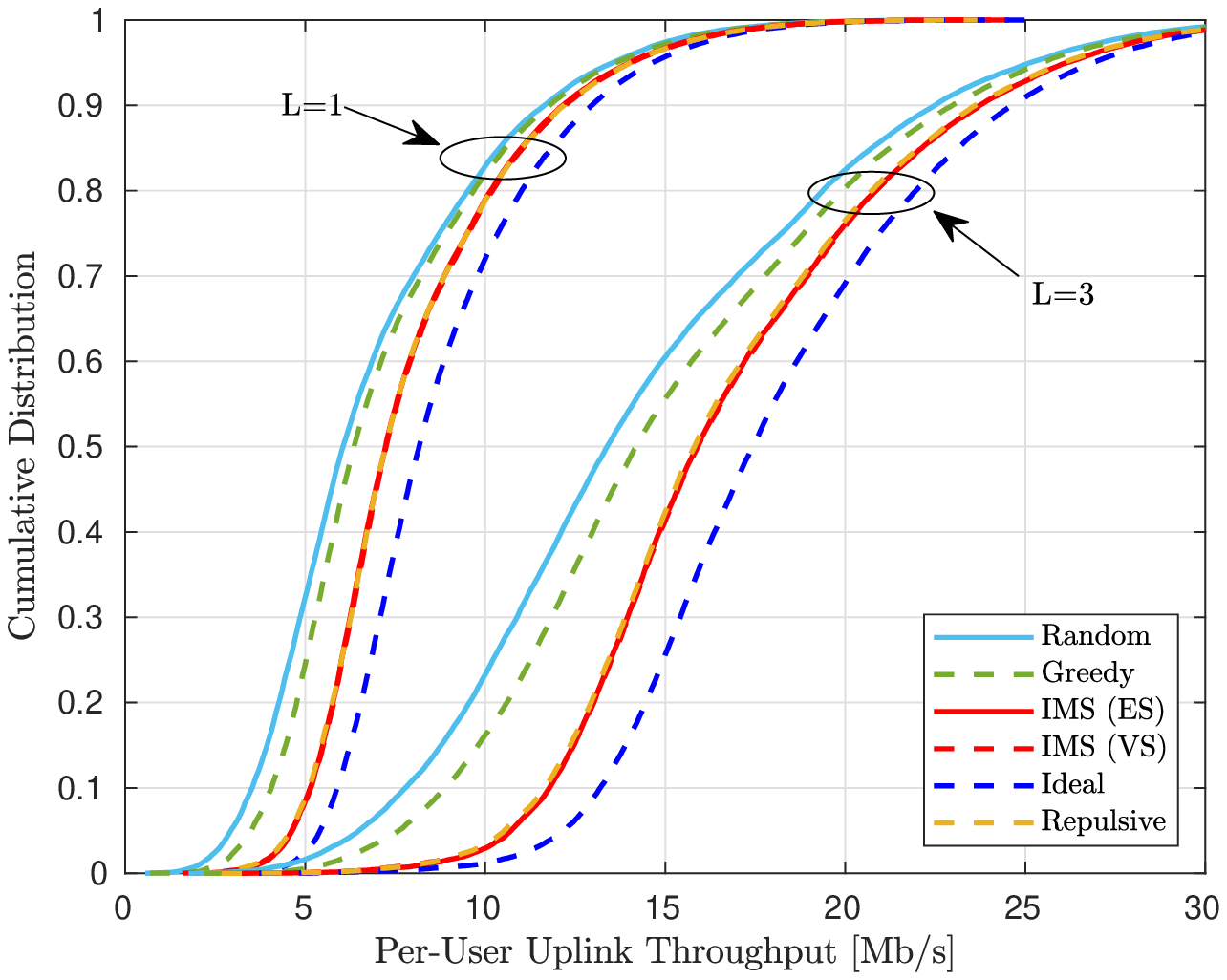}
         \label{fig:cdfUL}
     \end{subfigure}
     \hfill
     \begin{subfigure}[b]{0.53\textwidth}
         \centering
         \includegraphics[width=\textwidth]{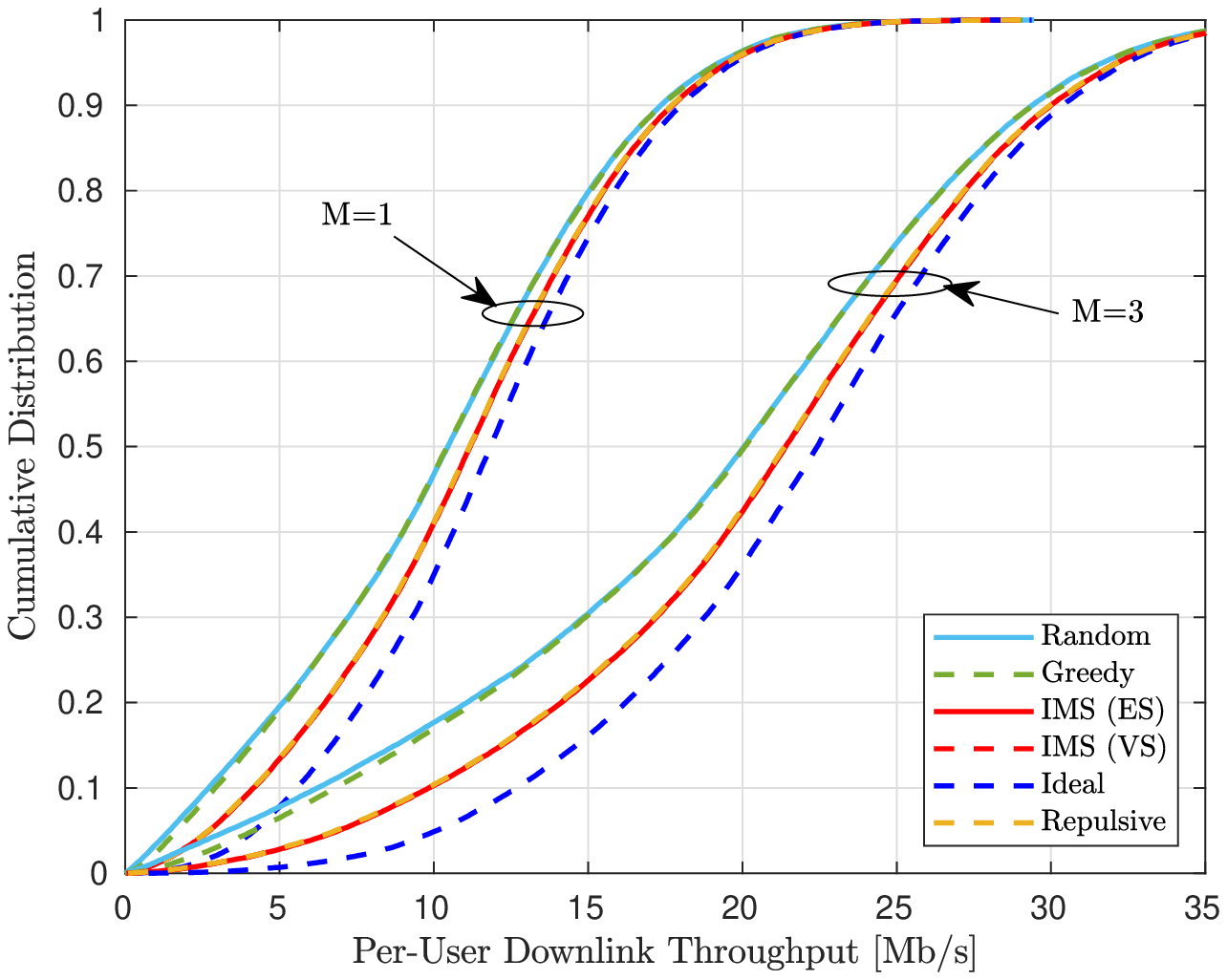}
         \label{fig:cdfDL}
     \end{subfigure}
        \caption{Cumulative distribution of the per-user uplink and downlink throughput for different pilot assignment strategies, $K=40$, $\tau_p=10$, and $M=200$. }
        \label{fig:cdf_Ls}
\end{figure}

The cumulative distribution of the per-user uplink and downlink throughput of different pilot assignment strategies for different numbers of antennas $L$ is shown in \Cref{fig:cdf_Ls}.
The superiority of the proposed scheme against other approaches is evident from the figure.
It can also be seen from the figure that increasing the number of antennas per \gls{ap} will increase the rate in all schemes while the gap between the proposed scheme and other algorithms also increases, which is reasonable as our scheme generates less co-pilot interference by properly utilizing the available resources.

\begin{table}[]
\centering
    \caption{95th percentile of per-user uplink and downlink throughput for different values of $L$, when $K=40$, $\tau_p=10$, and $M=200$.}
\begin{tabular}{lllll}
\hline
\multirow{2}{*}{Approach} & \multicolumn{2}{c}{UL}                            & \multicolumn{2}{c}{DL}                            \\ \cline{2-5} 
                          & \multicolumn{1}{c}{L=1} & \multicolumn{1}{c}{L=3} & \multicolumn{1}{c}{L=1} & \multicolumn{1}{c}{L=3} \\ \hline
Random                    & 2.93                    & 6.49                    & 1.43                    & 3.31                    \\
Greedy                    & 3.49                    & 7.59                    & 1.68                    & 4.16                    \\
Repulsive                 & 4.59                    & 10.52                   & 2.69                    & 6.80                    \\
IMS (ES)                   & 4.62                    & 10.71                   & 2.73                    & 6.90                    \\
IMS (VS)                   & 4.63                    & 10.68                   & 2.72                    & 6.88                    \\
Ideal                     & 5.40                    & 12.20                   & 4.18                    & 10.09                    \\ \hline
\end{tabular}\label{tbl:95thpercentile}
\end{table}


\begin{figure}
     \centering
     \begin{subfigure}[b]{0.53\textwidth}
         \centering
         \includegraphics[width=\textwidth]{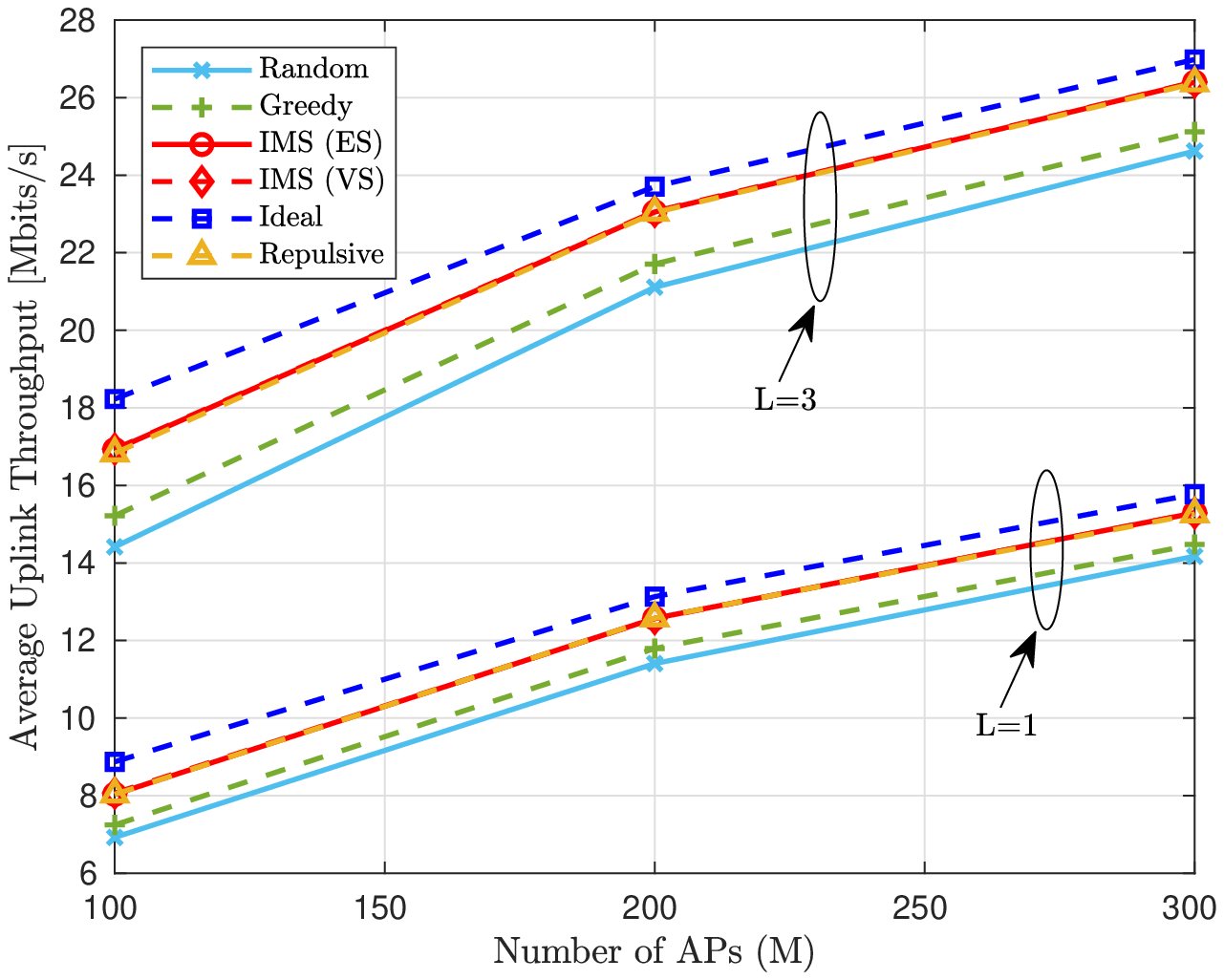}
         \label{fig:cdfUL}
     \end{subfigure}
     \hfill
     \begin{subfigure}[b]{0.53\textwidth}
         \centering
         \includegraphics[width=\textwidth]{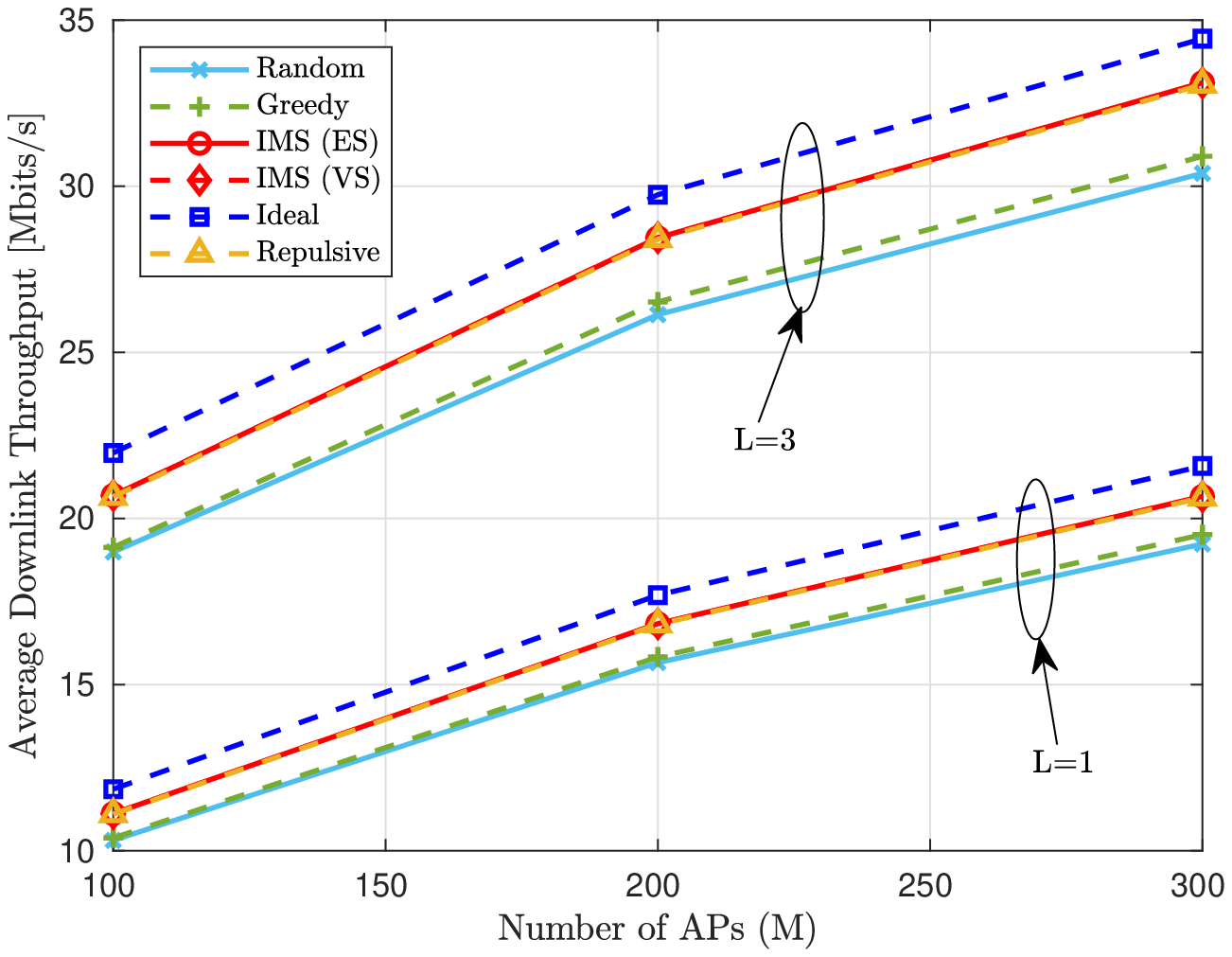}
         \label{fig:cdfDL}
     \end{subfigure}
        \caption{Average uplink and downlink throughput for different pilot assignment strategies, $K=40$, $\tau_p=10$. }
        \label{fig:diffM}
\end{figure}

\Cref{tbl:95thpercentile} shows the 95th percentile of the per-user throughput extracted from \Cref{fig:cdf_Ls}, where it can be seen that increasing the number of \glspl{ap}' antennas from $L=1$ to $L=3$ increases the 95th percentile by 2.3x for uplink and 2.5x for downlink.
Compared to other schemes, our approach performs slightly better than \textit{repulsive} clustering, but it improves the 95th percentile rate by 1.13~Mbps ($\approx32\%$) in uplink and 1.05~Mbps ($\approx 62\%$) in downlink over a \textit{greedy} pilot assignment scheme, when $L=1$.
These values for $L=3$ are 3.12~Mbps ($\approx38\%$) and 2.74~Mbps ($\approx 65\%$).

\begin{figure}
     \centering
     \begin{subfigure}[b]{0.53\textwidth}
         \centering
         \includegraphics[width=\textwidth]{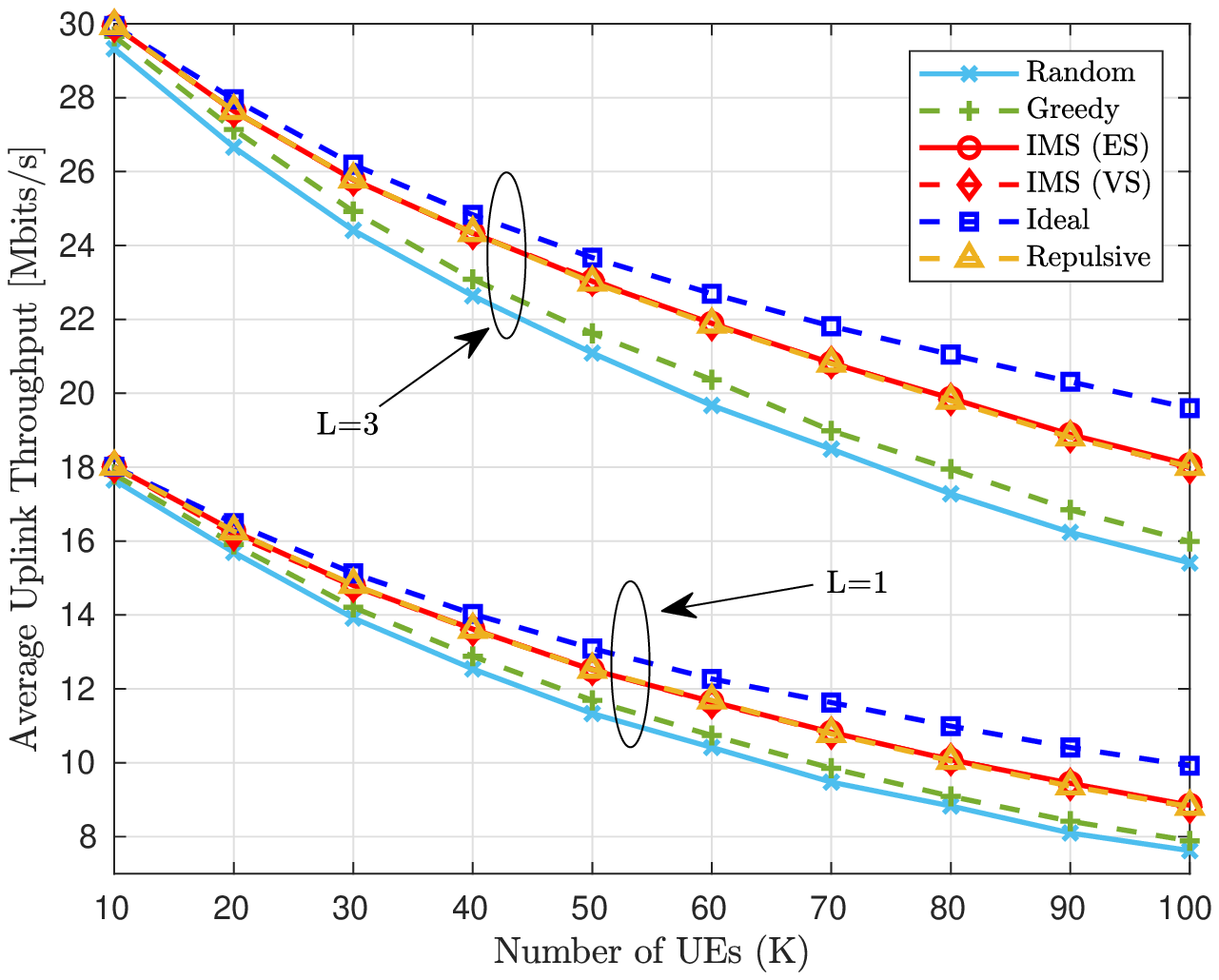}
         \label{fig:diffKUL}
     \end{subfigure}
     \hfill
     \begin{subfigure}[b]{0.53\textwidth}
         \centering
         \includegraphics[width=\textwidth]{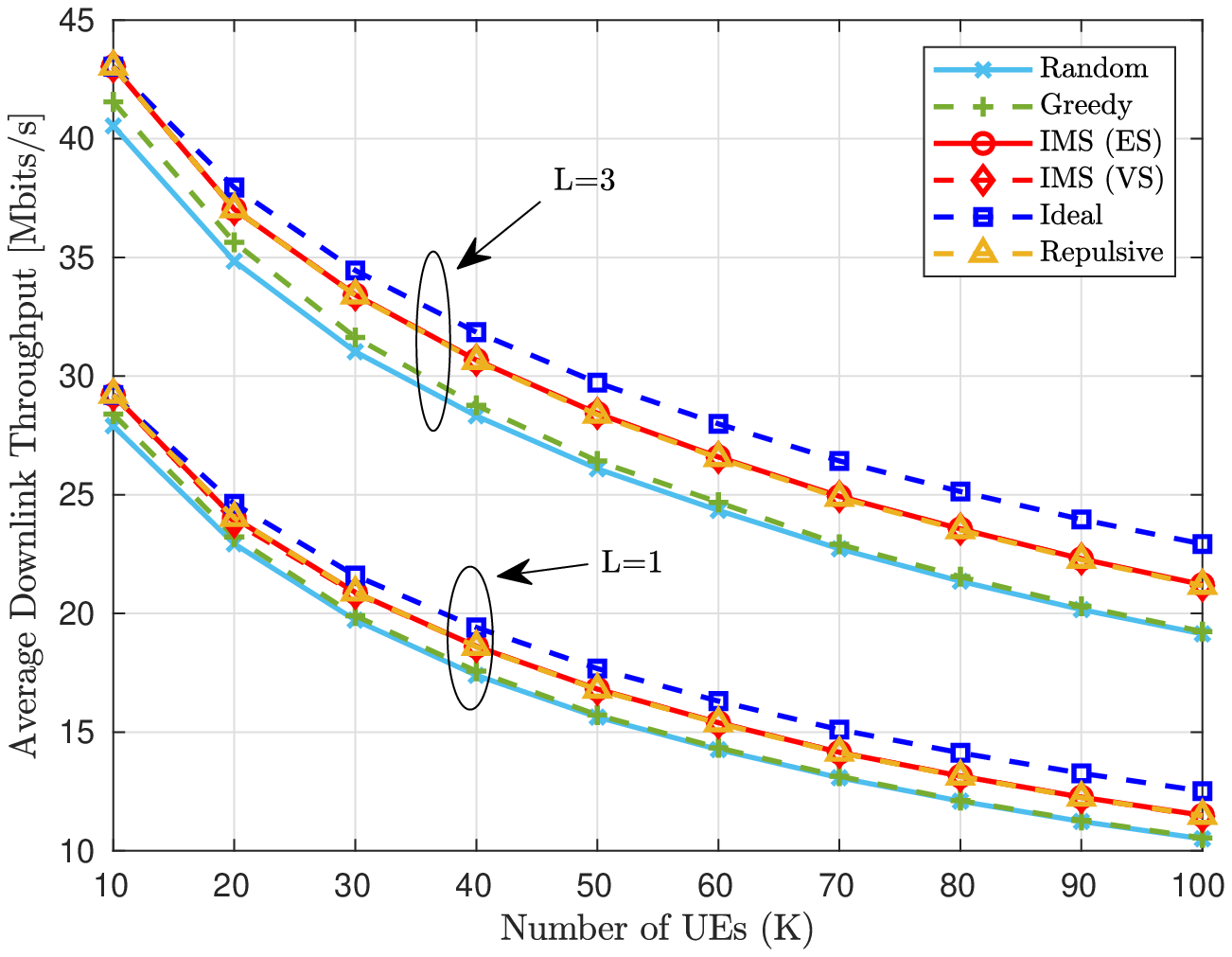}
         \label{fig:diffKDL}
     \end{subfigure}
        \caption{Average uplink throughput for different numbers of UEs, $M=100$ and $\tau_p=10$. }
        \label{fig:diffK}
\end{figure}
\Cref{fig:diffM} compares the average uplink and downlink throughput against different numbers of \glspl{ap} for different pilot reuse schemes.
By increasing the number of \glspl{ap}, the average throughput increases in both uplink and downlink.
Also, the performance of the multiple-antenna \glspl{ap} is always better than that of single-antenna \glspl{ap}.
For example, having 100 \glspl{ap} with three antennas (300 antennas in total) performs better than 300 single-antenna \glspl{ap}.
This comes from the fact that increasing $L$ makes the channel more favorable~\cite{chen2018channel} and reduces inter-user interference.
It also increases the array gain, which has already been discussed and analyzed in~\cite{mai2018cell}.

The average uplink and downlink throughput of different pilot assignment schemes against the number of \glspl{ue} is illustrated in~\Cref{fig:diffK}.
It can be seen from the figure that increasing the number of \glspl{ue} in the network, while the number of \glspl{ap} is fixed, will decrease the average throughput.
The throughput reduction ratio is different for the pilot assignment policies, and in our approach it is lower than in the \textit{random} and \textit{greedy} schemes.
This shows the reliability of our approach in large-scale scenarios.

\begin{figure}
     \centering
     \begin{subfigure}[b]{0.53\textwidth}
         \centering
         \includegraphics[width=\textwidth]{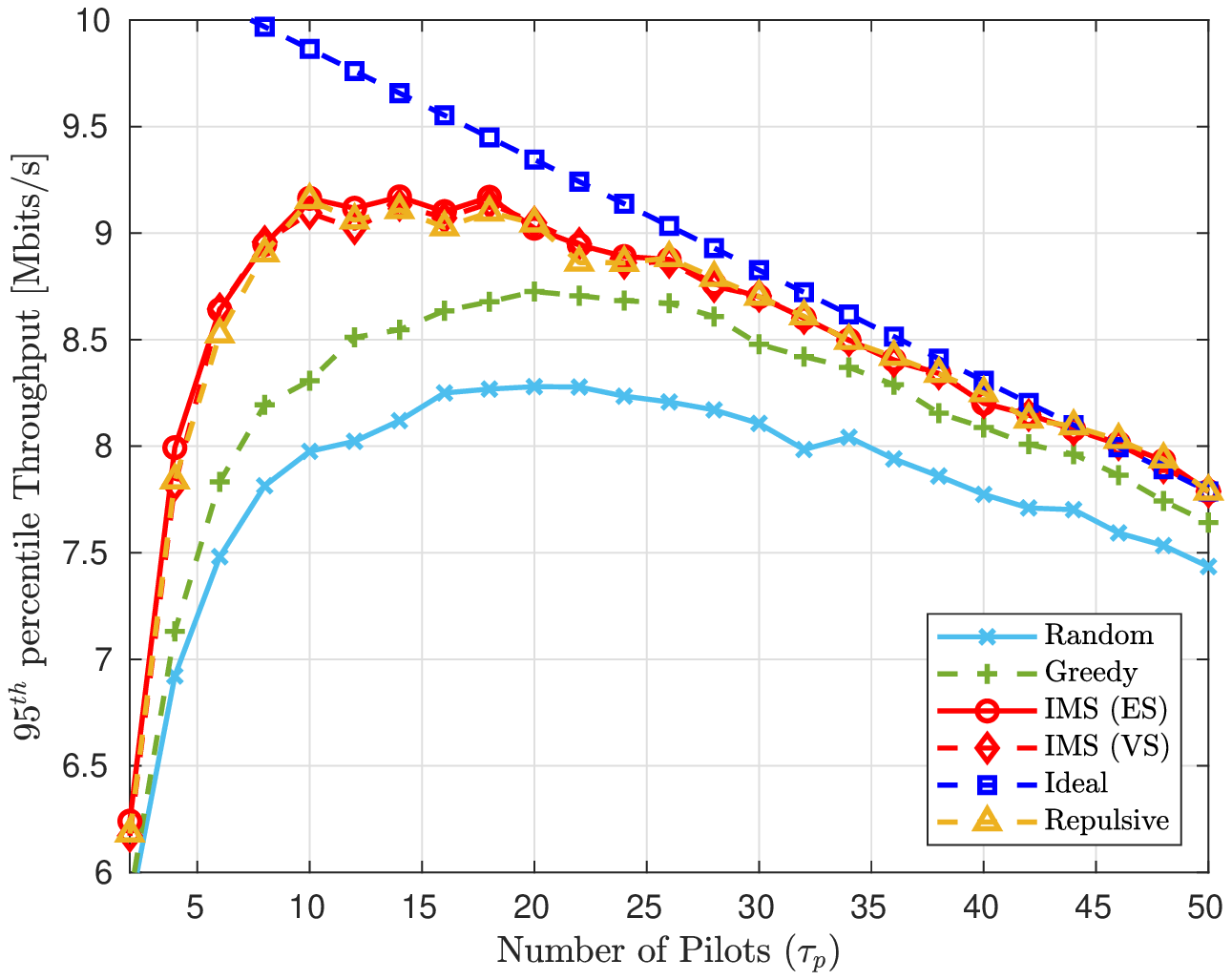}
         \label{fig:diffKUL}
     \end{subfigure}
     \hfill
     \begin{subfigure}[b]{0.53\textwidth}
         \centering
         \includegraphics[width=\textwidth]{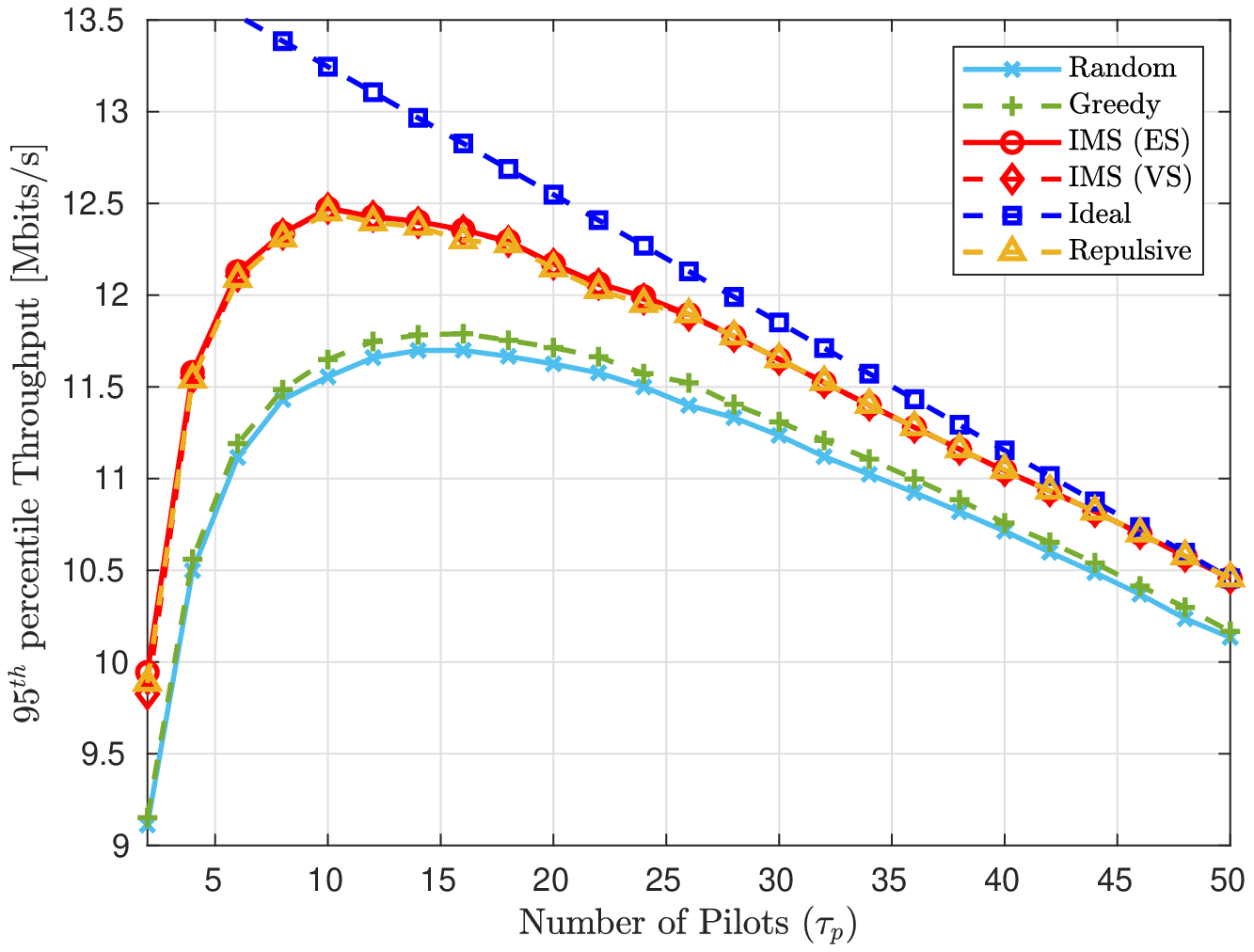}
         \label{fig:diffKDL}
     \end{subfigure}
        \caption{Average the per-user uplink and downlink throughput for different numbers of orthogonal pilots, $M=100$, $L=1$, and $K=50$. }
        \label{fig:diffTau}
\end{figure}

\Cref{fig:diffTau} presents the average uplink and downlink throughput with respect to the number of orthogonal pilots ($\tau_p$) for different pilot assignment schemes, while the size of the coherence block is fixed.
An interesting fact that can be seen in the figures is that increasing the number of pilots increases the performance only up to a certain point, after which the performance starts decreasing.
This shows the necessity of finding the optimal number of pilots, which is outside this paper's scope and is left for future research.

\begin{figure}[t]
     \centering
     \begin{subfigure}{0.53\textwidth}
         \centering
         \includegraphics[width=\textwidth]{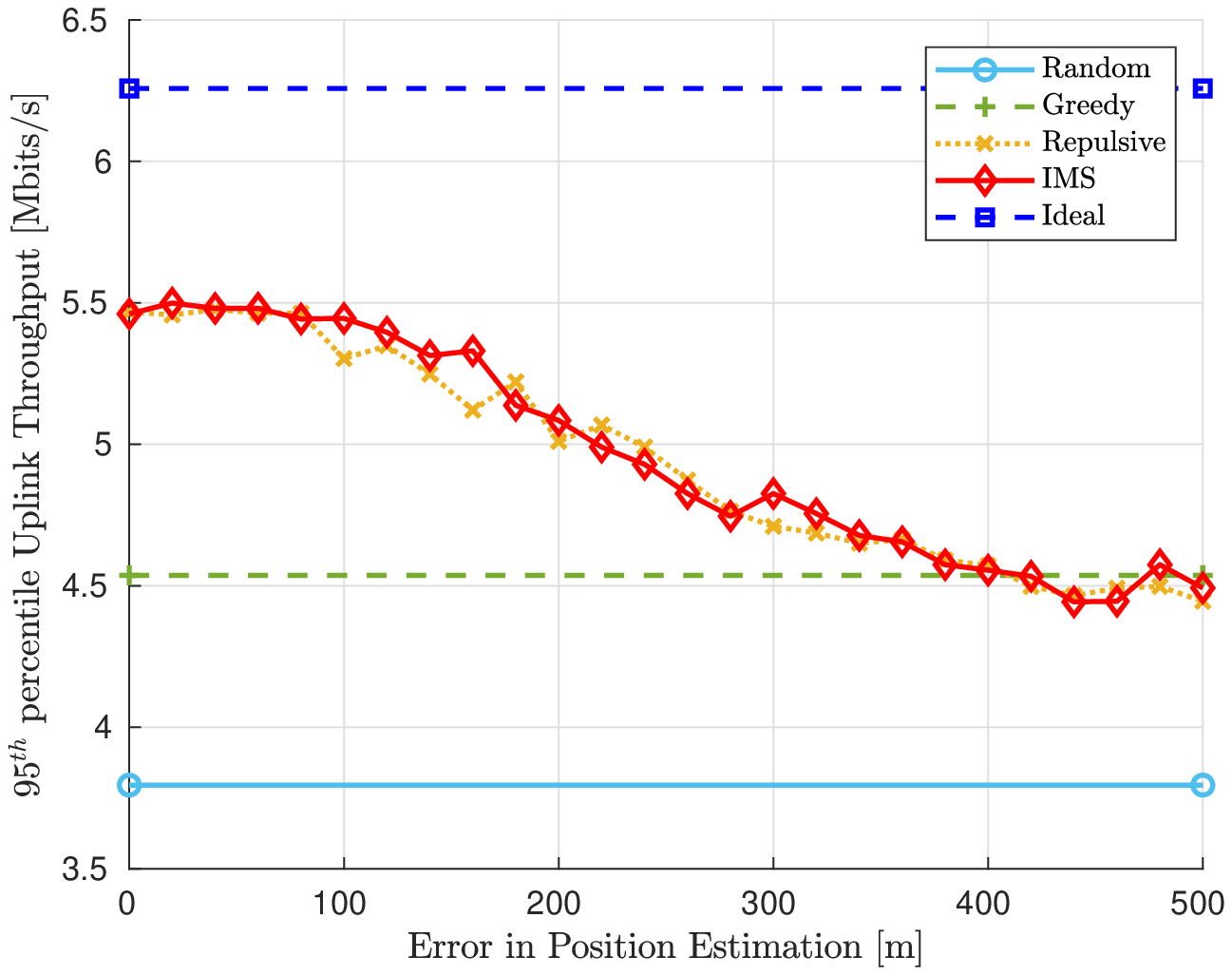}
         \label{fig:diffKUL}
     \end{subfigure}
     \begin{subfigure}{0.53\textwidth}
         \centering
         \includegraphics[width=\textwidth]{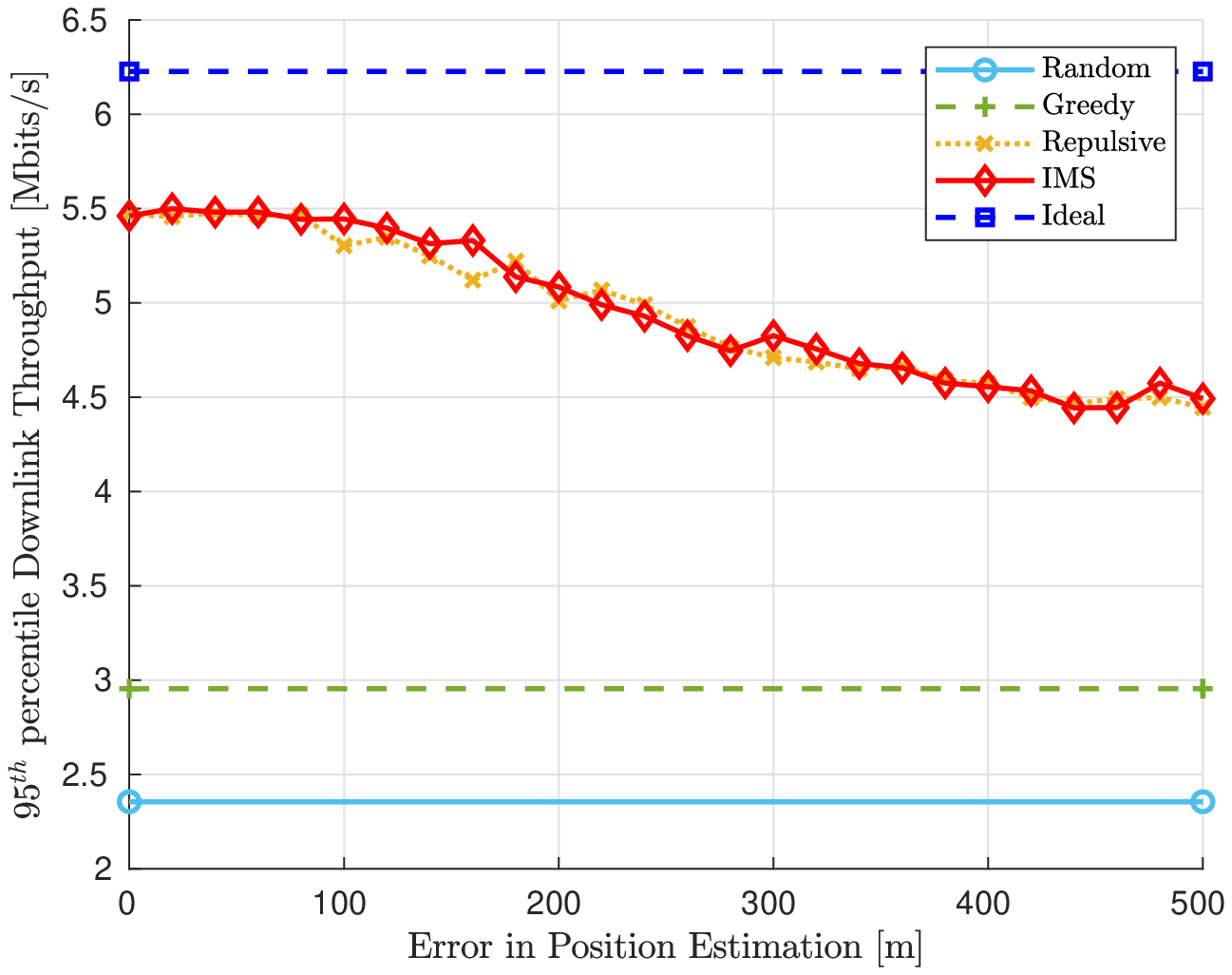}
         \label{fig:diffKDL}
     \end{subfigure}
        \caption{Average uplink and downlink throughput for different values of error in location estimation, $M=120$, $K=50$, $K=10$, and $L=1$.}
        \label{fig:locations}
\end{figure}

\Cref{fig:locations} shows the 95\textsuperscript{th} percentile rate of the proposed schemes in the presence of errors in the \glspl{ue} locations estimation.
As seen in the figure, the proposed algorithms, until a certain level, are robust against errors, and the system performance is not affected much when the error in the location estimation is less than 100 meters.
Even after that, the rate is higher than in \textit{random} and \textit{greedy} schemes.

\begin{figure}[t]
     \centering
     \begin{subfigure}{.53\textwidth}
         \centering
         \includegraphics[width=\textwidth]{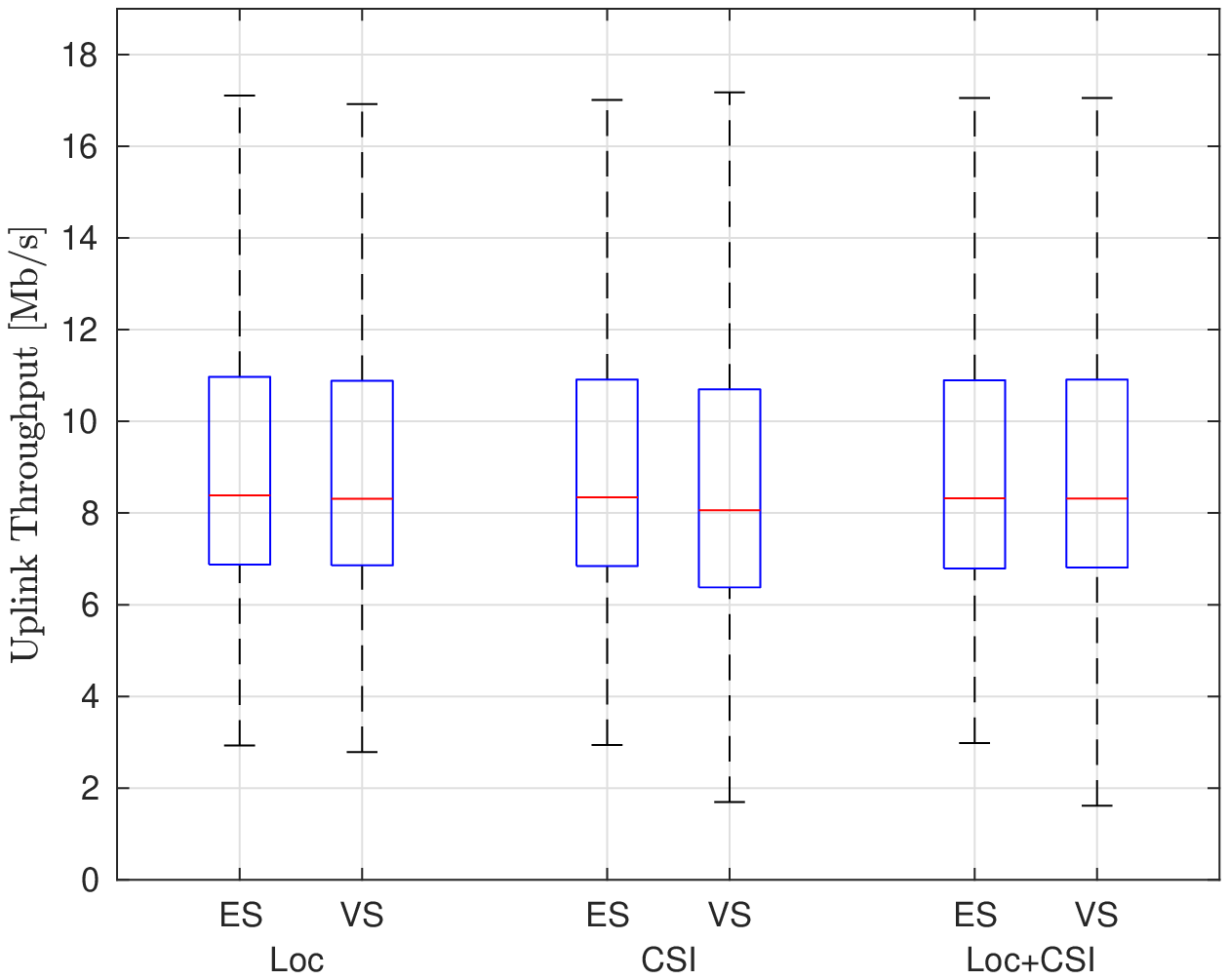}
         \label{fig:locVsCSIUL}
     \end{subfigure}
     \begin{subfigure}{.53\textwidth}
         \centering
         \includegraphics[width=\textwidth]{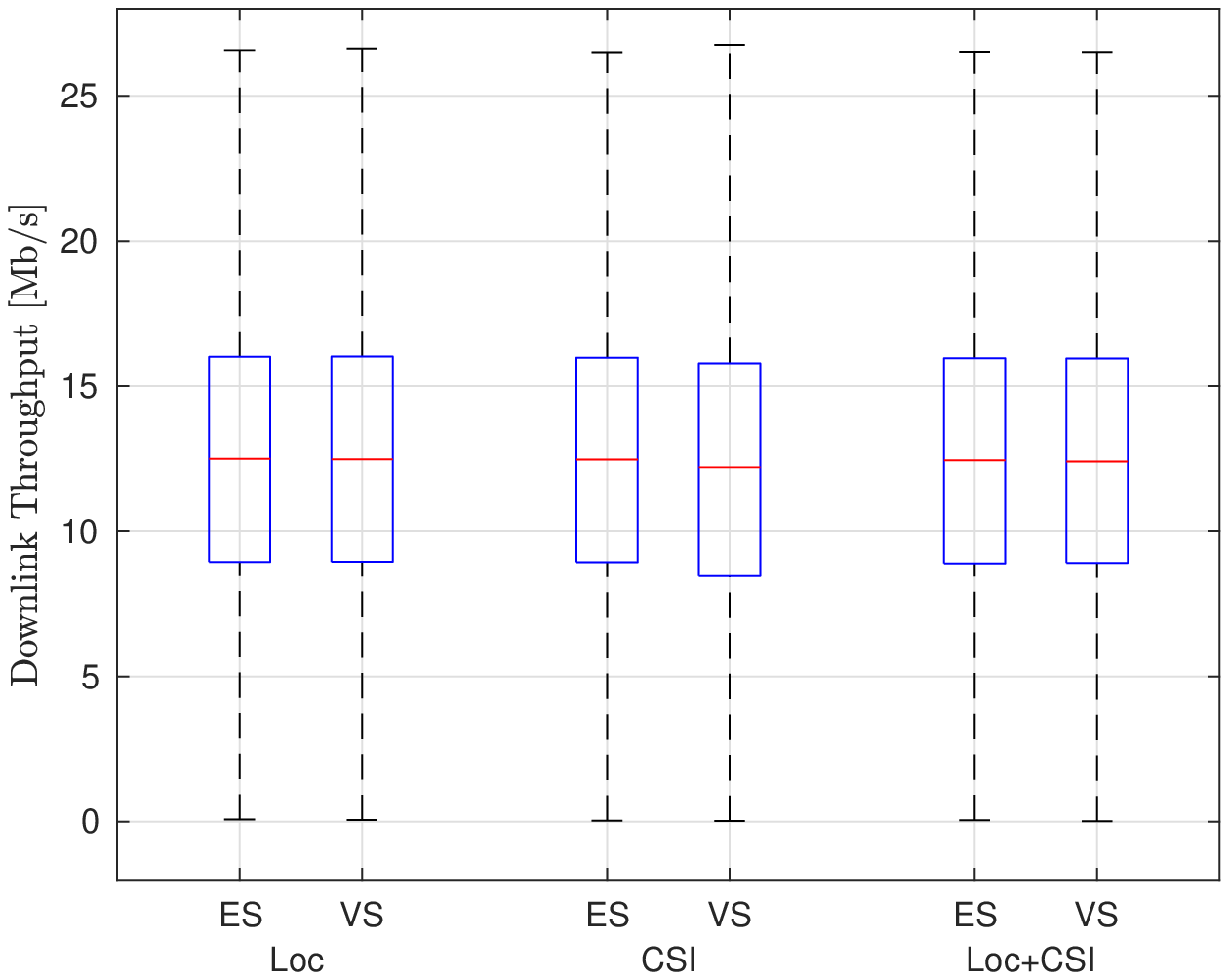}
         \label{fig:locVsCSIDL}
     \end{subfigure}
        \caption{Uplink and downlink throughput when location and/or CSI is used as input feature for clustering.}
        \label{fig:locationsVsCSI}
\end{figure}

\Cref{fig:locationsVsCSI} shows the uplink and downlink throughput for \glspl{ue} when the \gls{csi} (\gls{lsf}) and/or \gls{ue}'s locations are considered as input features in our clustering algorithm.
The figure shows that both features provide relatively similar results, so each can be used as input when the other feature's data is unavailable.
It can also be seen that combining both features does not provide better solutions, and considering only one feature space is enough.




\section{Conclusion}\label{sec:conclusion}
\gls{cf} \gls{mmimo} will be an essential part of future wireless communication systems, given its capability of providing uniform service for the \glspl{ue}.
The performance of these systems can still be further improved before it becomes a functional and operational technology.
Pilot contamination is recognized as an undesirable effect that can highly degrade the performance of \gls{cf} \gls{mmimo} when the \glspl{ue} share the same pilot.
In this paper, we formulated pilot assignment in \gls{cf} \gls{mmimo} systems as a \gls{dcp} problem and proposed an iterative maxima search approach to solve it.
Numerical results show the proposed scheme's effectiveness compared to other approaches from the literature.
In future works, we will expand our approach by replacing the Euclidean distance with more sophisticated and parameterized diversity functions, i.e., \glspl{dnn} that consider different networking factors such as \gls{ap} locations, and the density of \glspl{ue} and \glspl{ap}.
Another extension will consider pilot assignment jointly with pilot power control, which can further improve the channel estimation performance.
The scalability of different pilot assignment strategies is another factor that should be considered in future research.


\ifCLASSOPTIONcaptionsoff
  \newpage
\fi

\bibliographystyle{IEEEtran}
\bibliography{journal_v03_arxiv.bib}



\end{document}